\documentclass[lettersize,journal]{IEEEtran}

\usepackage{graphicx} 
\usepackage{epstopdf}
\usepackage{subfigure}
\usepackage{stfloats}
\usepackage{algorithm, algorithmic}
\usepackage{diagbox}
\usepackage{multirow}
\usepackage{mathtools}
\usepackage{setspace}  
\usepackage{threeparttable}
\usepackage{textcomp,booktabs}
\usepackage[usenames,dvipsnames]{color}
\usepackage{colortbl}
\usepackage{indentfirst}
\usepackage{cite}
\usepackage{amsmath,amssymb,amsfonts}
\usepackage{algorithmic}
\usepackage{textcomp}
\usepackage{xcolor}
\usepackage{mathtools}
\usepackage{amsthm}
\usepackage{color}
\usepackage{enumitem}
\usepackage[hang,flushmargin]{footmisc}

\newtheorem{lemma}{\bf Lemma}
\newtheorem{proposition}{\bf Proposition}

\theoremstyle{definition}

\definecolor{mygray}{gray}{.9}
\definecolor{mypink}{rgb}{.99,.91,.95}
\definecolor{mycyan}{cmyk}{.3,0,0,0}

\begin{document}
\title{A General Optimization Framework for Tackling Distance Constraints in Movable Antenna-Aided Systems}
\author{Yichen Jin, Qingfeng Lin, Yang Li, Hancheng Zhu, Bingyang Cheng, Yik-Chung Wu, \textit{Senior Member, IEEE}, and Rui Zhang, \textit{Fellow, IEEE} 
\thanks{\quad An earlier version of this paper was presented in part at the IEEE SPAWC 2024~\cite{Yichen}.}
\thanks{\quad Y. Jin, H. Zhu, B. Cheng, and Y.-C. Wu are with the Department of Electrical and Electronic Engineering, The University of Hong Kong, Hong
Kong (e-mail: u3589542@eee.hku.hk, u3006551@connect.hku.hk, bycheng@eee.hku.hk, ycwu@eee.hku.hk).}
\thanks{\quad Q. Lin is with the Department of Electrical and Electronic Engineering, The University of Hong Kong, Hong
Kong, and also with Shenzhen Research Institute of Big Data, Shenzhen 518172, China (e-mail: qflin@eee.hku.hk).}
\thanks{\quad Y. Li is with the School of Computing and Information Technology, Great Bay University, Dongguan 523000, China, and also with Shenzhen Research Institute of Big Data, Shenzhen 518172, China (e-mail: liyangblade@gmail.com).}
\thanks{\quad R. Zhang is with the School of Science and Engineering, Shenzhen
Research Institute of Big Data, The Chinese University of Hong Kong, Shenzhen, Guangdong 518172, China (e-mail: rzhang@cuhk.edu.cn), and also with the Department of
Electrical and Computer Engineering, National University of Singapore, Singapore 117583 (e-mail: elezhang@nus.edu.sg).}
}

\maketitle

\begin{abstract}

The recently emerged movable antenna (MA) shows great promise in leveraging spatial degrees of freedom to enhance the performance of wireless systems. However, resource allocation in MA-aided systems faces challenges due to the non-convex and coupled constraints on antenna positions. This paper systematically reveals the challenges posed by the minimum antenna separation distance constraints. Furthermore, we propose a penalty optimization framework for resource allocation under such new constraints for MA-aided systems. Specifically, the proposed framework separates the non-convex and coupled antenna distance constraints from the movable region constraints by introducing auxiliary variables. Subsequently, the resulting problem is efficiently solved by alternating optimization, where the optimization of the original variables resembles that in conventional resource allocation problem while the optimization with respect to the auxiliary variables is achieved in closed-form solutions. To illustrate the effectiveness of the proposed framework, we present three case studies: capacity maximization, latency minimization, and regularized zero-forcing precoding. Simulation results demonstrate that the proposed optimization framework consistently outperforms state-of-the-art schemes.

\end{abstract}

\begin{IEEEkeywords}
Distance separation constraints, movable antenna, non-convex and coupled constraints, penalty optimization, resource allocation.
\end{IEEEkeywords}

\section{Introduction} \label{intro}


The recently emerged communication applications, including 4K video streaming and augmented reality/virtual reality, have become an integral part of future wireless systems. These applications have raised the demand for higher wireless network capacity, but improving transmission rates using conventional antenna technologies becomes challenging because the currently available degrees of freedom (DoF) have been exhausted~\cite{Historical_Review_FA_MA}. To expand the DoF in a cost-effective manner, the movable antenna (MA), or the fluid antenna (FA), has emerged as a promising solution~\cite{Historical_Review_FA_MA, MPA_MA_wirelesscomm, fast_FAMA_Massive_connect}. 


Compared to conventional antenna systems, MA/FA systems leverage additional DoF in wireless systems by adjusting the antenna positions. In particular,~\cite{FAS} advocates using antennas made of fluid materials, so that the antennas could move freely within a specified area, searching for better communication channel with higher signal-to-noise ratio (SNR). Alternatively, by connecting a large number of closely spaced prefabricated antennas to the radio-frequency (RF) chain via flexible cables, the MA positions can be flexibly adjusted in real-time via stepper motors~\cite{MA_WC_oppo_and_chal}. Furthermore, six-dimensional MA (6DMA) has been recently proposed as a general model with three-dimensional (3D) position and 3D rotation adjustments of MAs~\cite{zhangrui1,zhangrui2,zhangrui3}. Given the demonstrated benefits of incorporating MA/FA/6DMA into existing wireless systems across diverse scenarios~\cite{MIMO_cap_cha_for_MA, Dynamic_Beam_Coverage_Satellite_MA, Efficient_Algorithm_SumRateMaxi_FAS_ISAC, IRS-aided_wirelesscomm_ME, MUComm_MA_BS_via_APO, FAS_MUMIMO_ISAC_DRL}, they are expected to be a strong contender for enhancing system performance in 6G and beyond~\cite{6D_MA_user_distribution_M&O}.    


However, to prevent antenna coupling effects~\cite{MIMO_cap_cha_for_MA}, the MA introduces new and intricate constraints related to the minimum antenna separation distance. Consequently, the feasible set of MA positions is obtained by eliminating some parts of the antenna panel that do not meet the antenna distance constraint. This makes the feasible set of MA positions more complex than constraints in conventional wireless resource allocation problems. Worse still, as the non-convex antenna distance constraints are coupled together, it is impossible to handle these constraints separately.

Existing attempts for handling these unconventional MA-induced constraints and the associated nontrivial feasible set can be broadly classified into three categories. 

\subsubsection{The zeroth-order algorithms~\cite{FA_MEC, MA_wireless_powered_MEC, MA_empoweredPLS_withoutCSI_JO_B&AP, HA2,HA6}} This approach uses algorithms like particle swarm optimization (PSO) and simulated annealing (SA) to address the problem. Nonetheless, these algorithms rely on random search and typically require heuristic parameter tuning. 

\subsubsection{Simplifying the feasible set through conservative approximations~\cite{MA_Multiusercomm_APO, MA_secure_trans_without_eaveCSI, GD0_FA_MUuplink_lowcomplexity_GD,GD8,GD9}} In this approach, the feasible set of MAs is divided into multiple conservative non-overlapping subsets, ensuring that any two points within these subsets satisfy the distance constraints~\cite{MIMO_cap_cha_for_MA,MA_secure_trans_without_eaveCSI}. While this approach allows for the simultaneous updating of all MA positions, the reduction of the feasible set results in a performance loss. This loss is particularly pronounced when dealing with a large number of MAs or a small antenna panel. Furthermore, if there are a large number of MAs, it may be infeasible to identify a group of non-overlapping subsets with a simple topology.


\subsubsection{Updating the position of each MA sequentially~\cite{MIMO_cap_cha_for_MA, Flexible_precoding_MU_MAcomm, Efficient_Algorithm_SumRateMaxi_FAS_ISAC, Dynamic_Beam_Coverage_Satellite_MA, SCA1,SCA4,SCA8,SCA9,SCA10,SCA11,SCA14,SCA15,GD1,GD7}} This sequential approach involves optimizing the position of one MA at a time while keeping the positions of other MAs fixed. This is by far the most popular approach in the literature. However, the sequential updating of MA positions often results in the partitioning of the feasible set into multiple unconnected subsets. This makes typical iterative algorithms, such as the successive convex approximation (SCA) and gradient-based methods, get trapped in one of the subsets without exploring other unconnected subsets.

This paper aims to address the challenges of managing the new coupled and non-convex distance constraints induced by MAs. In particular, this paper, for the first time, presents a comprehensive analysis of the challenges arising from MA-induced constraints and their implications on existing methods for resource allocation. Then, we propose a penalty optimization framework, which provides a flexible and generalized paradigm for optimizing wireless resources in MA-aided systems. To avoid the unconnected subsets appearing in the sequential update of MA positions, the antenna distance constraints are separated from other wireless resource constraints by introducing auxiliary variables and penalty terms to the objective function. With the optimization of MA positions together with other wireless resources being separated from that of the auxiliary variables as two subproblems, the antenna distance constraints only appear in the latter. By optimizing the two subproblems iteratively while increasing the penalty parameter in each iteration~\cite{qingfeng}, the final solution gradually satisfies the MAs' minimum separation distance constraints. 

In the proposed penalty optimization framework, the subproblem with respect to the original variables becomes a resource allocation problem without MA-induced antenna distance constraints, therefore can be solved by a wide variety of methods, including conventional optimization methods and the more modern deep learning methods. On the other hand, for the subproblem with respect to auxiliary variables, the challenge for solving it persists due to the retention of non-convex and coupled constraints. Despite the challenge, this paper derives a new closed-form solution for this subproblem.

The proposed framework has many advantages compared to existing methods. Firstly, the proposed framework does not resort to conservative approximation of the feasible set, without performance loss caused by the reduction of the feasible set and is applicable to any number of MAs. Secondly, although the framework still involves sequential updates, there are no unconnected subsets due to the introduced auxiliary variables. As a result, the proposed framework exhibits less reliance on the initial variable values. Thirdly, the proposed framework is practical and user-friendly as it requires no tuning of hyperparameters, making it straightforward to implement. The above advantages faciliate seamless integration of MA with other wireless technologies, and pave the way for the future incorporation of artificial intelligence (AI)-based algorithms.

To demonstrate the effectiveness of the proposed framework, three examples of MA-aided system optimization problem are illustrated: capacity maximization, latency minimization, and regularized zero-forcing (RZF) precoding. Simulation results demonstrate the convergence of the proposed framework and show it consistently outperforms state-of-the-art approaches in all evaluated scenarios. These examples serve as concrete guidance for adopting the proposed framework in a broad range of MA-related optimization problems.

Compared to its previous conference version~\cite{Yichen}, this paper provides new and thorough analysis of the constraints in MA-aided systems and reveals for the first time the corresponding challenges from an optimization perspective. For the proposed framework, this paper also provides an in-depth discussion on the solution of the subproblem with respect to the auxiliary variables. Moreover, this paper delves into a wider range of MA-related resource allocation problems to showcase the versatility and robustness of the proposed method.

The remainder of this paper is organized as follows. Section~\ref{New Constraints} presents the new constraints in MA-aided systems and their associated challenges. In Section~\ref{framework}, we propose a penalty optimization framework for the resource allocation in MA-aided systems. Section~\ref{case} provides three case studies to demonstrate the effectiveness of the proposed framework in different scenarios. Section~\ref{Numerical Results} presents numerical results and discussions under various setups. Finally, Section~\ref{Conclusion} concludes the paper.

\textit{Notations:} Unless otherwise specified, the vectors involved in this paper are column vectors. Symbols for vectors (lower case) and matrices (upper case) are presented in boldface. The transpose and conjugate transpose of a vector or matrix are denoted by $(\cdot)^T$ and $(\cdot)^H$, respectively. The symbol Tr($\mathbf{A}$) represents the trace of matrix $\mathbf{A}$, and diag($\mathbf{a}$) denotes a square diagonal matrix with the elements of the vector $\mathbf{a}$ placed on the main diagonal. Moreover, $\mathbf{A} \succeq 0$ indicates that matrix $\mathbf{A}$ is positive semi-definite, and $\mathcal{CN}(\mathbf{0},\mathbf{\Sigma})$ refers to the circularly symmetric complex Gaussian distribution with zero mean and covariance matrix $\mathbf{\Sigma}$. The identity matrix of order $K$ is represented by $\mathbf{I}_K$. The 2-norm of vector $\mathbf{a}$ is denoted as $||\mathbf{a}||_2$, while $||\mathbf{A}||_2$ and $||\mathbf{A}||_F$ represent the spectral norm and Frobenius norm of the matrix $\mathbf{A}$, respectively. The expectation is denoted by $\mathbb{E}(\cdot)$. The squiggle letters, such as $\mathcal{C}$, $\mathcal{W}$ and $\mathcal{U}$, refer to sets. The real part of a complex number is denoted by $\Re\{\cdot \}$, and $\circ$ denotes the Hadamard product.

\section{Challenges due to MA-induced Constraints}   \label{New Constraints}

Let $\text{MA}_m$ denote the $m$-th ($m = 1, 2, \dots, M$) MA at the antenna panel, and $\mathbf{r}_m \triangleq [x_m, y_m]^T$ denote the position of $\text{MA}_m$. To avoid antenna coupling, the MA-aided systems generally require the following constraints~\cite{MIMO_cap_cha_for_MA}:
\begin{align} \label{given_region}
\mathbf{r}_m \in \mathcal{C}, ~\forall m =1,2, \ldots, M, 
\end{align}
\begin{align} \label{antenna_distance}
 \left\|\mathbf{r}_m-\mathbf{r}_l\right\|_2 \geq D, ~  \forall m,l=1,2, \ldots, M, ~ m \neq l,
\end{align}
where $\mathcal{C}$ represents the allowable region for the MAs to move, which is commonly a convex set, and $D$ denotes the minimum distance between each pair of antennas.

\begin{figure}
    \centering \includegraphics[width=0.65\linewidth]{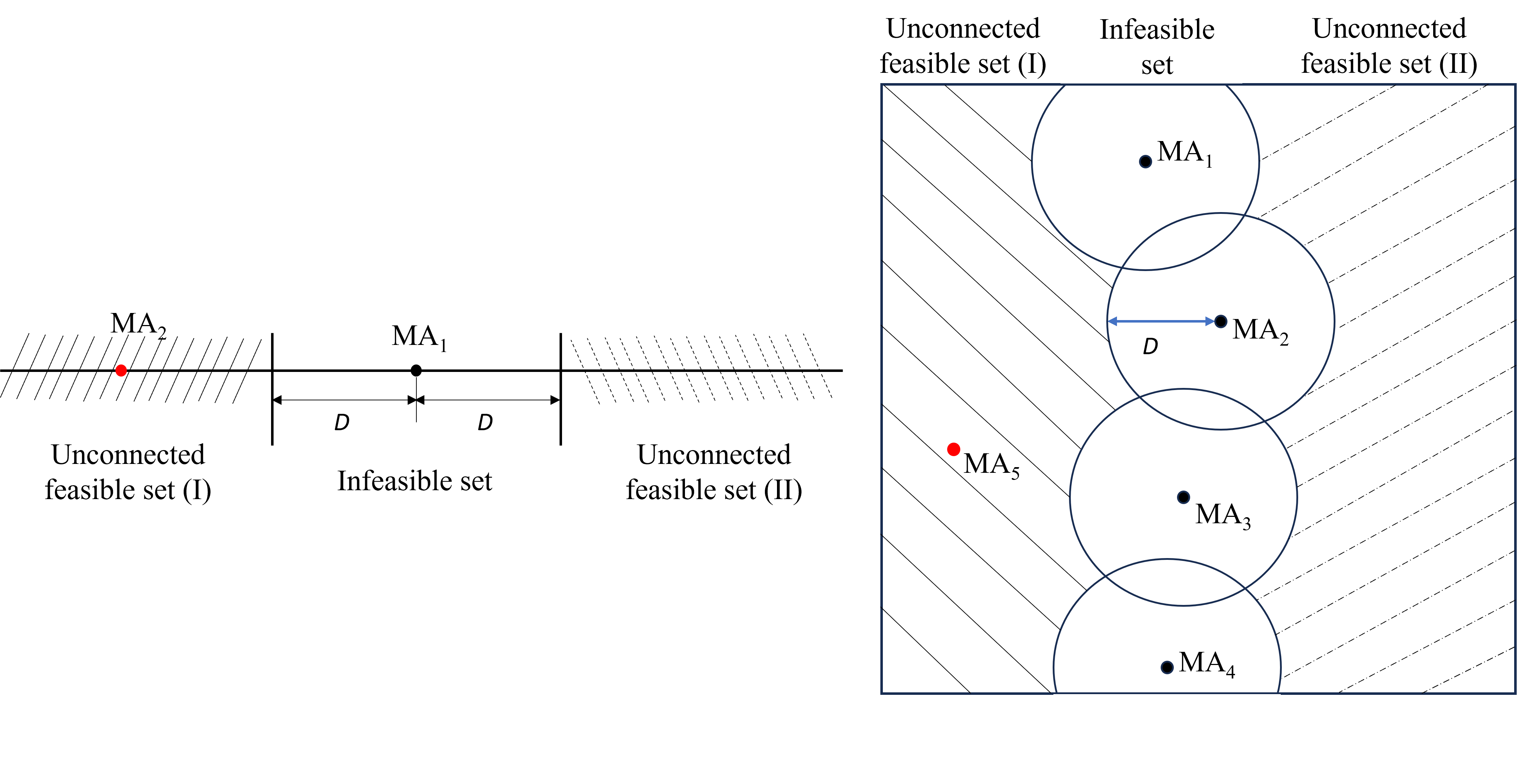}
    \caption{Illustration of unconnected feasible sets (with respect to the `red' MA) with sequential MA position optimization: 1D case (left) and 2D case (right).}
    \label{fig:sequential}
\end{figure}

It is apparent that the feasible MA positions must satisfy both the constraints on the given region and antenna distance. Since the antenna panel area is finite, this may result in the feasible set being partitioned into several unconnected subsets. This is particular problematic for optimization methods that sequentially update the positions of MA, as in~\cite{MIMO_cap_cha_for_MA,SCA4,SCA7,SCA8,SCA10}. For example, as shown in Fig.~\ref{fig:sequential}, the feasible sets for the position of $\text{MA}_2$ (for 1D case) and $\text{MA}_5$ (for 2D case) are two unconnected regions when the other antenna positions are fixed. 
Any iterative algorithm starting from one feasible subset could not jump to another one. Worse still, when the number of MA increases, the feasible set might even be partitioned into more than two unconnected subsets, making the optimization even more challenging.

\begin{figure}
   \centering \includegraphics[width=0.65\linewidth]{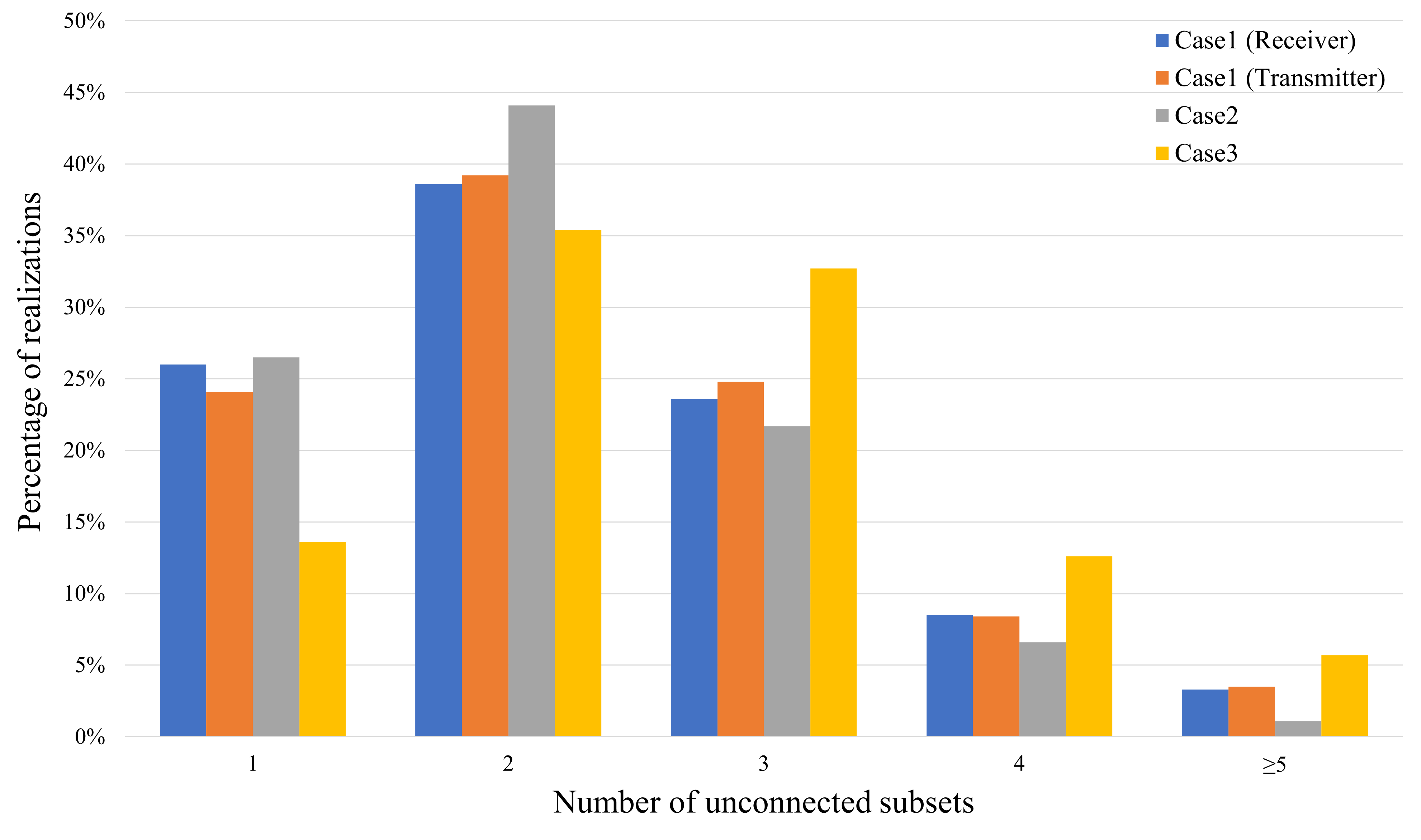}
   \caption{Chances of the antenna panel being divided into unconnected feasible sets in three case studies in Section~\ref{case}.}
   \label{fig:unconnected}
\end{figure}

To further illustrate the prevalent phenomenon of unconnected feasible sets in MA-related resource allocation problems, Fig.~\ref{fig:unconnected} depicts the number of unconnected subsets in the antenna panel for all the case studies in Section~\ref{case}. In particular, the system consists of $6$ MAs, in a 2D antenna panel with each side twice that of the carrier wavelength, and $D$ equals half a wavelength. The data in Fig.~\ref{fig:unconnected} is obtained from $1000$ channel realizations, and the sequential MA position optimization is run for $3$ iterations. From Fig.~\ref{fig:unconnected}, it can be seen that there are about $70$ to $85\%$ of the channel realizations being impacted by unconnected feasible set. 

\begin{figure}
    \centering \includegraphics[width=0.65\linewidth]{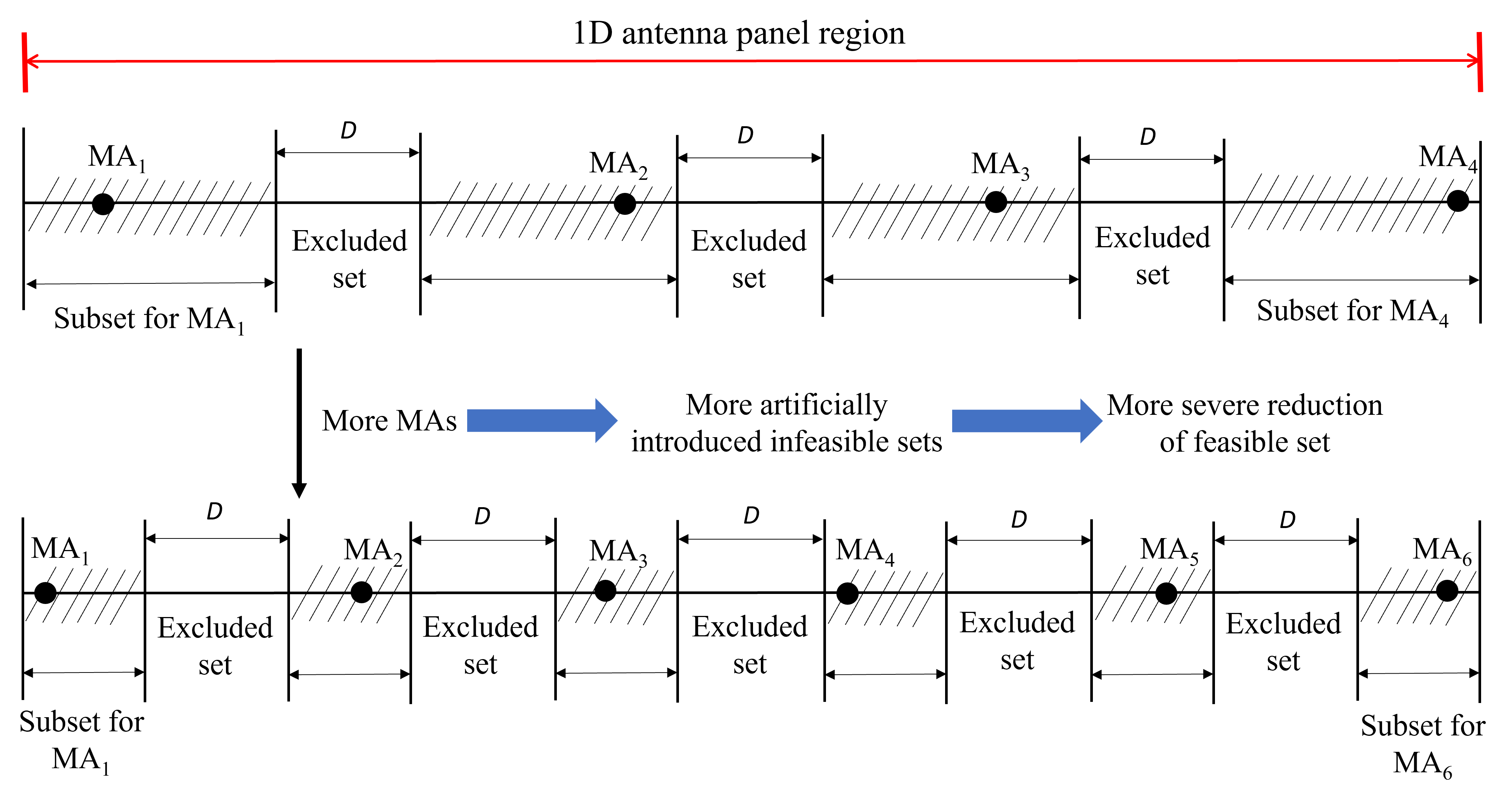}
    \caption{Illustration of the conservative approximation of the feasible set.}
    \label{fig:conservative}
\end{figure}

On the other hand, conservative approximation might avoid unconnected feasible subsets but it also results in severe performance loss caused by the reduction of the feasible set. More specifically, in a 1D MA system, the MA position constraints in the original problem are replaced by a group of pre-assigned box constraints before optimization ~\cite{MA_secure_trans_without_eaveCSI}. An example is shown in Fig.~\ref{fig:conservative} where certain regions in a 1D antenna panel (marked as excluded set in Fig.~\ref{fig:conservative}) are excluded such that the MAs can freely move within their respective feasible subsets while without violating the antenna distance constraints. However, this approach leads to a severe reduction of the feasible set. For example, assume that $\text{MA}_1$ is optimized at the position depicted in Fig.~\ref{fig:conservative}. In order to maximize the utilization of the feasible set for other MAs, the excluded region with length $D$ should start immediately to the right of the $\text{MA}_1$ position, rather than the pre-assigned excluded sets as shown in Fig.~\ref{fig:conservative}. Therefore, pre-fixing the feasible subset for each MA before optimization is a conservative approach.

When the number of MAs is large, conservative approximation of the feasible set might not even be workable. For example, if the 1D MA array has an antenna panel length of $3D$, and we approximate the feasible set with line segment subsets of length $D$, then the maximum number of antennas that the panel can accommodate is only $2$. In general, for a 1D antenna panel of length $A$ and subsets of length $B$, the maximum number of MAs that can be handled through conservative approximation is given by $\lfloor \frac{A+D}{B+D} \rfloor$. Similarly, for a 2D antenna panel of size $A_{1} \times A_{2}$ and subsets of size $B_{1} \times B_{2}$, the maximum number of MAs is $\lfloor \frac{A_{1}+D}{B_{1}+D} \rfloor \times \lfloor \frac{A_{2}+D}{B_{2}+D} \rfloor$.

In the next section, we will propose a penalty optimization framework that overcomes the above challenges in the MA-aided systems.

\section{A Penalty Optimization Framework for MA-Aided Systems}   \label{framework}

Consider the following optimization problem:
\begin{align}
\mathcal{P}: &\min_{\{\mathbf{r}_m\}_{m=1}^M, \mathbf{X}} ~ f\left(\{\mathbf{r}_m\}_{m=1}^M, \mathbf{X}\right) \nonumber \\
&~~~~~\text { s.t. } \quad
 \mathbf{X} \in \mathcal{X}, ~\eqref{given_region}, \text{ and}~\eqref{antenna_distance}, \nonumber
\end{align}
where $f(\cdot)$ denotes a general utility function, $\mathbf{X}$ denotes the variables other than the MA positions with its feasible set denoted by $\mathcal{X}$. For example, in the MA-aided multiple-input multiple-output (MIMO) systems of~\cite{MIMO_cap_cha_for_MA}, $f(\cdot)$ denotes the system capacity, and $\mathbf{X}$ is the transmit covariance matrix.

In general, if $f(\cdot)$ is a non-convex objective function or $\mathcal{X}$ is a non-convex set, the optimization for $\mathcal{P}$ is nontrivial even without the MAs. Furthermore, the coupling between $\{\mathbf{r}_m\}_{m=1}^M$ and $\mathbf{X}$ in the objective function together with the non-convex constraints~\eqref{antenna_distance} add another layer of challenges. A standard approach is to alternately optimize $\mathbf{X}$ and $\{\mathbf{r}_m\}_{m=1}^M$ as two blocks of variables~\cite{liyang1}. In this approach, the subproblem for optimizing $\mathbf{X}$ resembles the traditional resource allocation problem. However, the subproblem for optimizing $\{\mathbf{r}_m\}_{m=1}^M$ is challenging due to the coupled non-convex antenna distance constraints~\eqref{antenna_distance}. As discussed in the previous section, common approaches to handle the coupled non-convex constraints~\eqref{antenna_distance} involve either sequentially updating each MA position or conservative approximation to simplify the constraints. But these two approaches may both lead to significant performance loss caused by either the disconnection or the reduction of the feasible set.

\subsection{The Penalty Optimization Framework}

To avoid the unconnected feasible sets appearing in the optimization, we propose to use variable splitting by introducing $\{\mathbf{z}_m = \mathbf{r}_m\}_{m=1}^M$. By replacing $\{\mathbf{r}_m\}_{m=1}^M$ in~\eqref{antenna_distance} with $\{\mathbf{z}_m\}_{m=1}^M$ and adding a penalty term $\rho \sum_{m=1}^M \left\|\mathbf{r}_m - \mathbf{z}_m\right\|_2^2$ to the objective function, $\mathcal{P}$ becomes $\mathcal{P}1$ on the top of this page. In $\mathcal{P}1$, $\rho>0$ is the penalty factor and it is known that when $\rho$ goes to infinity, the optimal solutions of $\mathcal{P}$ and $\mathcal{P}1$ are identical~\cite{penalty_goes_to_inf}. By resorting to auxiliary variables, the proposed framework separates the antenna distance constraint from the allowable region constraint. This ensures that there are no unconnected subsets during optimization, thus avoiding performance loss caused by getting stuck in one subset without exploring other unconnected subsets.

\begin{figure*}
\begin{align}
\mathcal{P}1:&\min_{\{\mathbf{r}_m, \mathbf{z}_m\}_{m=1}^M, \mathbf{X}} f\left(\{\mathbf{r}_m\}_{m=1}^M, \mathbf{X}\right) + \rho \sum_{m=1}^M \left\|\mathbf{r}_m - \mathbf{z}_m\right\|_2^2 \label{ori_p}\\
&~~~~~~~\text { s.t. } 
~~~~~\quad \mathbf{X} \in \mathcal{X}, \nonumber \\
&~~~~~~~~~~~~~~~~~~~~~ \mathbf{r}_m \in \mathcal{C}, ~~\forall m =1,2, \ldots, M, \nonumber \\
&~~~~~~~~~~~~~~~~~\quad \left\|\mathbf{z}_m-\mathbf{z}_l\right\|_2 \geq D,  ~~~~\forall m,l=1,2, \ldots, M, ~~ m \neq l. \nonumber
\end{align}
\hrulefill
\end{figure*}

The problem $\mathcal{P}1$ can be solved by alternating optimization (AO) with respect to the original variable of $\mathcal{P}$ and $\{\mathbf{z}_m\}_{m=1}^M$. The details of each subproblem are given in the following.


\subsubsection{Subproblem with respect to $\mathbf{X}$ and $\{\mathbf{r}_m\}_{m=1}^M$} This subproblem is given by
\begin{align}
\mathcal{P}\text{1-a}:~~&\min_{\mathbf{X}, \{\mathbf{r}_m\}_{m=1}^M} ~~ f\left(\{\mathbf{r}_m\}_{m=1}^M, \mathbf{X}\right) + \rho \sum_{m=1}^M \left\|\mathbf{r}_m - \mathbf{z}_m\right\|_2^2 \nonumber \\
&~~~~\text { s.t. } \quad \mathbf{X} \in \mathcal{X} \nonumber \\ 
&~~~~\quad\quad\quad \mathbf{r}_m \in \mathcal{C}, ~\forall m =1,2, \ldots, M. \nonumber
\end{align}
Note that subproblem $\mathcal{P}\text{1-a}$ only involves the original variable of $\mathcal{P}$ without the most intricate antenna distance constraints, and thus it can be handled by many existing methods. For instance, optimizing with respect to $\mathbf{X}$ can be addressed using existing approaches designed for conventional fixed-position antennas (FPA) systems. On the other hand, when optimizing $\{\mathbf{r}_m\}_{m=1}^M$, the simple geometry of the antenna panel often results in $\mathbf{r}_m \in \mathcal{C}$ being a box constraint, which can be effectively managed by various methods, such as the projected gradient method.

\subsubsection{Subproblem with respect to $\{\mathbf{z}_m\}_{m=1}^M$}          Since the auxiliary variables $\{\mathbf{z}_m\}_{m=1}^M$ only appear in the penalty term, the resulting subproblem is given by
\begin{align} 
\mathcal{P}\text{1-b}:~~&\min_{\{\mathbf{z}_m\}_{m=1}^M} \sum_{m=1}^M \left\|\mathbf{z}_m - \mathbf{r}_m\right\|_2^2 \nonumber \\
&~~~~\text { s.t. } ~~\left\|\mathbf{z}_m-\mathbf{z}_l\right\|_2 \geq D,  \nonumber \\ &~~~~~~~~~~~~\forall m,l=1,2, \ldots, M, ~~ m \neq l. \nonumber
\end{align}
Although $\mathcal{P}\text{1-b}$ remains non-convex and its optimal solution is generally difficult to obtain, the original distance constraint~\eqref{antenna_distance} regarding the MA position variables $\{ \mathbf{r}_m \}_{m=1}^M$ is re-expressed in terms of axillary variables $\{\mathbf{z}_m\}_{m=1}^M$. This approach allows the MA position variables $\{ \mathbf{r}_m \}_{m=1}^M$ to look for a solution without the distance constraints in $\mathcal{P}$ during updates~\cite{unified_framework_STAR-RIS}, preventing the reduction of the feasible set of $\{\mathbf{r}_m\}_{m=1}^M$, even when the $\mathbf{z}_m$ for each $m$ is sequentially updated. When $\{\mathbf{z}_l\}_{l\neq m}$ is fixed, the $m$-th subproblem of $\mathcal{P}\text{1-b}$ is given by
\begin{align} \label{p1_b-m}
\mathcal{P}\text{1-b-m}:~~&\min_{{\mathbf{z}_m}}   \left\|\mathbf{z}_m - \mathbf{r}_m\right\|_2^2 
 \nonumber \\
&~\text { s.t. } 
\left\|\mathbf{z}_m-\mathbf{z}_l\right\|_2 \geq D, \nonumber \\
&\quad \quad \forall l=1,2, \ldots, M, \quad l \neq m. \nonumber
\end{align}

The proposed framework involves  iteratively solving $\mathcal{P}\text{1-a}$ and $\mathcal{P}\text{1-b}$ (which sequentially solves $\mathcal{P}\text{1-b-m}$ for different $m$). After each outer iteration between $\mathcal{P}\text{1-a}$ and $\mathcal{P}\text{1-b}$, the penalty factor $\rho$ is increased by a factor larger than $1$. The whole procedure is summarized in Algorithm~\ref{a1}. With regard to the stopping criterion, it can be determined by the value of $\rho$, the number of iteration, or the relative change of objective value. For each fixed $\rho$, Algorithm~\ref{a1} ensures the updates for all subproblems result in monotonically non-increasing objective function value of $\mathcal{P}\text{1}$ in each outer iteration. Moreover, if the gradient of the objective function of $\mathcal{P}$ remains bounded, as $\rho$ increases, $\mathbf{r}_m$ will gradually converge to $\mathbf{z}_m$, satisfying the distance constraint~\eqref{antenna_distance}.

\begin{algorithm}[t] 
\caption{The Penalty Framework for Solving $\mathcal{P}$ } 
\begin{algorithmic}[1] \label{a1}
\STATE Initialize the optimization variables
\REPEAT
\STATE Optimize $\mathbf{X}$ and $\{\mathbf{r}_m\}_{m=1}^M$ in $\mathcal{P}\text{1-a}$.
\STATE \textbf{for} $m=1,\ldots,M$ \textbf{do}
\STATE ~~~~Optimize $\mathbf{z}_m$ in $\mathcal{P}\text{1-b-m}$.
\STATE \textbf{end for}
\STATE Increase the penalty factor $\rho$.
\UNTIL Stopping criterion is satisfied.	
\end{algorithmic}
\end{algorithm}

\subsection{The Optimal Solution of $\mathcal{P}\text{1-b-m}$}
It can be observed that $\mathcal{P}\text{1-b-m}$ essentially involves finding a $\mathbf{z}_m$ that minimizes its distance from $\mathbf{r}_m$, while also ensuring that the distances between $\mathbf{z}_m$ and other $\{\mathbf{z}_l\}_{l\neq m}$ are all larger than or equal to $D$.
\begin{figure*}  
	\centering
    \subfigure[$|\mathcal{L}|=0$]{\label{111} 
		\includegraphics[width=1.2in]{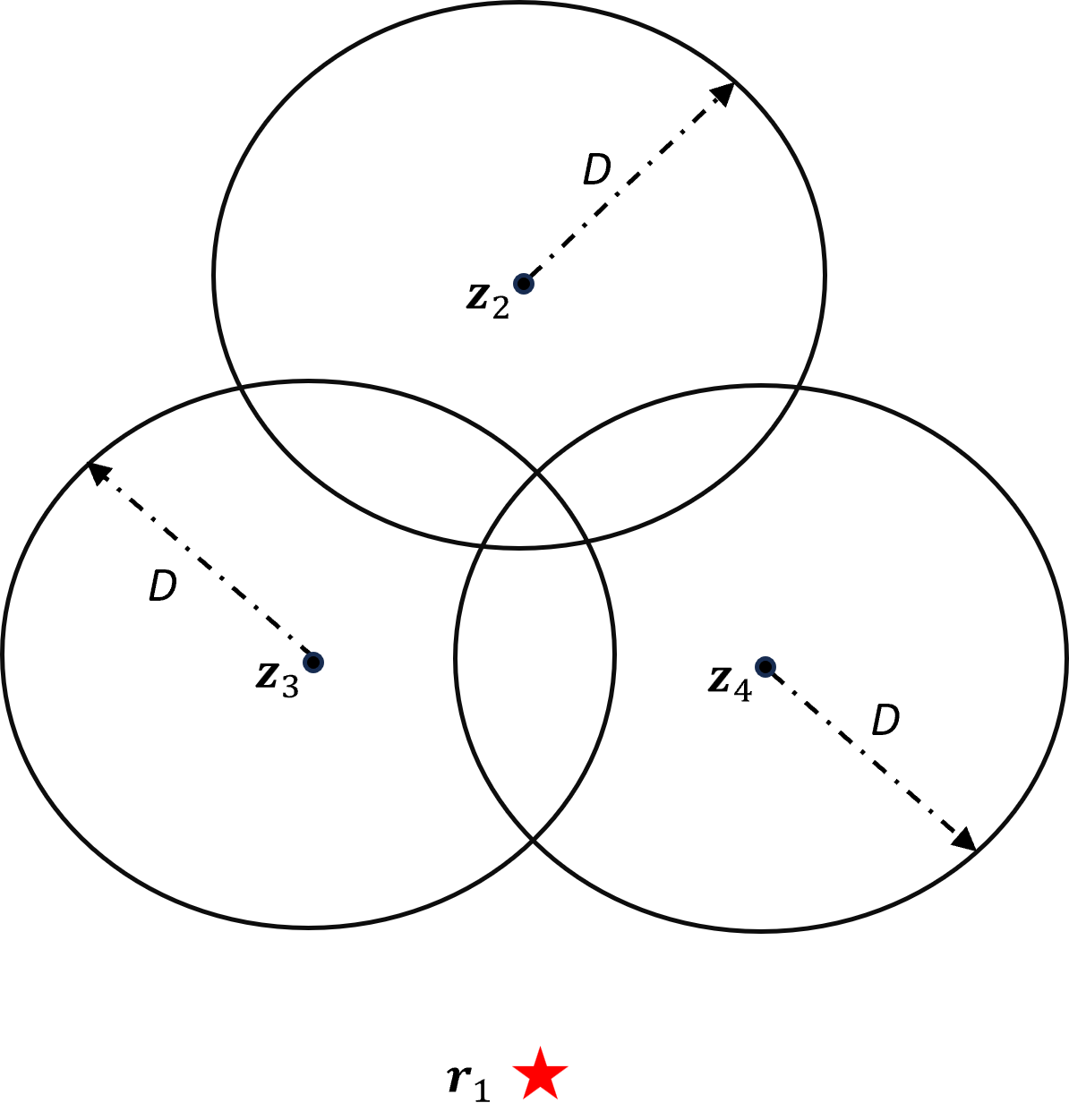}} 
    \subfigure[$|\mathcal{L}|=1$]{\label{222} 
		\includegraphics[width=1.2in]{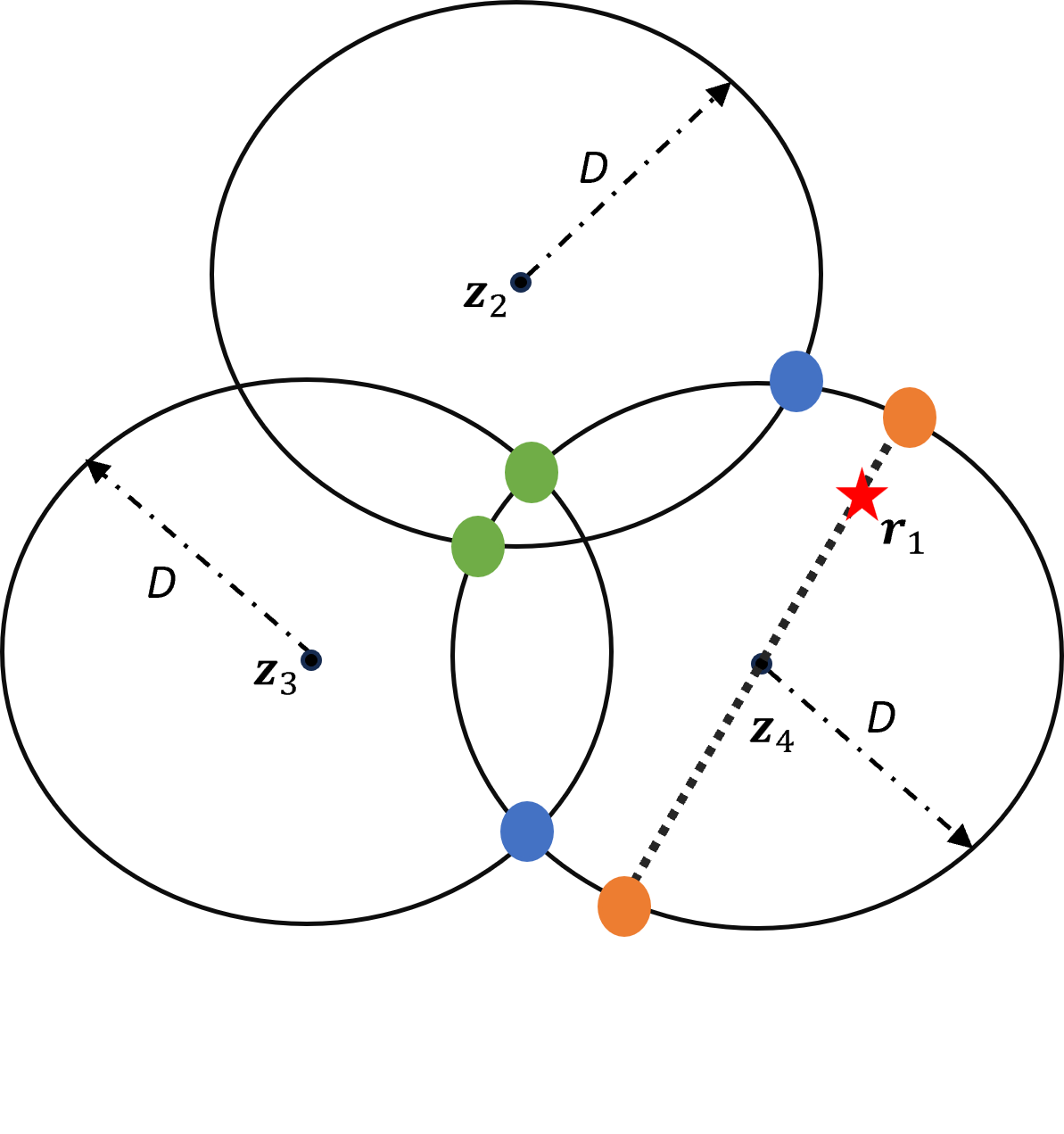}} 
	\subfigure[$|\mathcal{L}|=1$]{\label{333} 
		\includegraphics[width=1.2in]{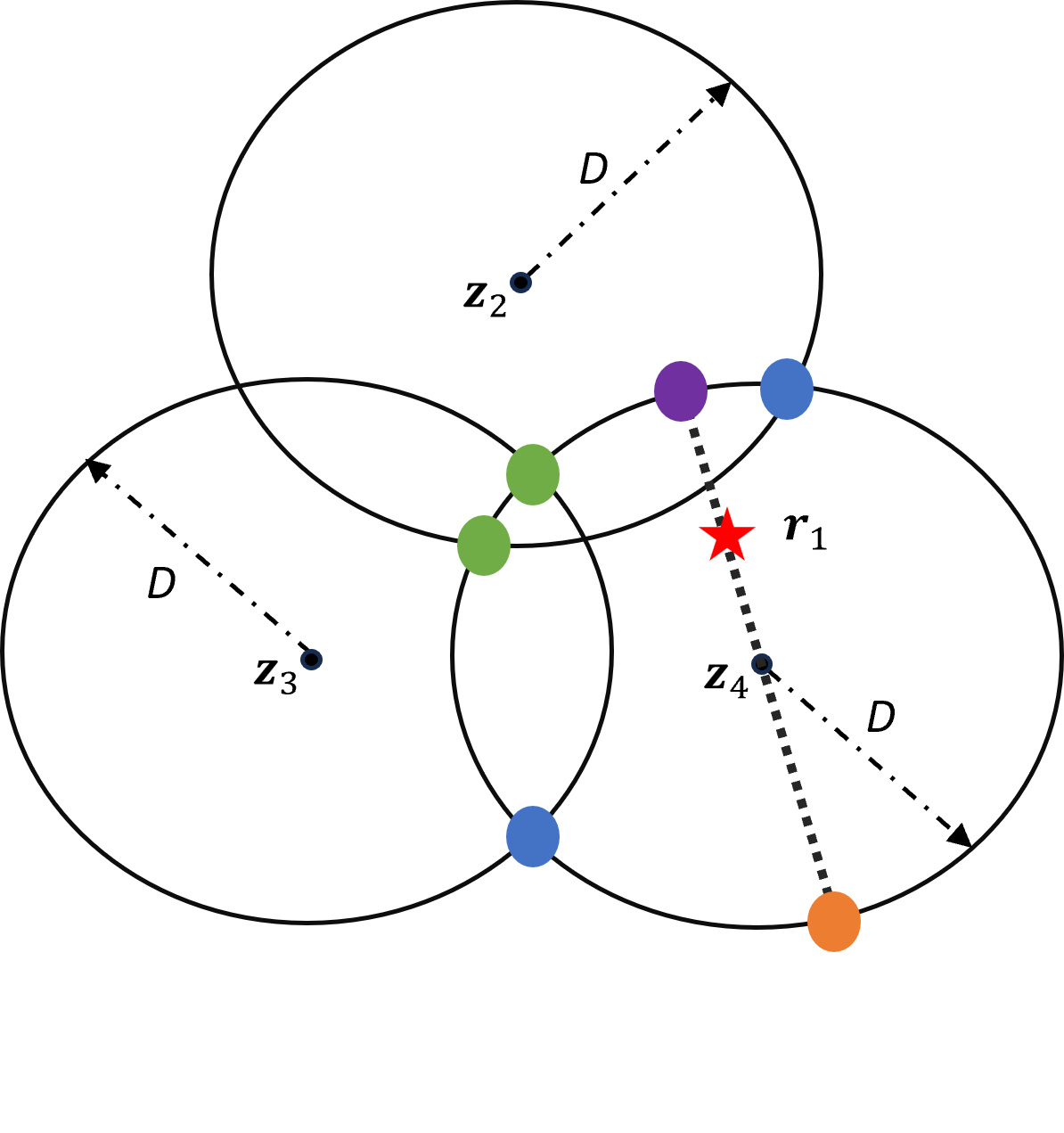}}  
	\subfigure[$|\mathcal{L}|=2$ ]{
		\label{666} 
		\includegraphics[width=1.2in]{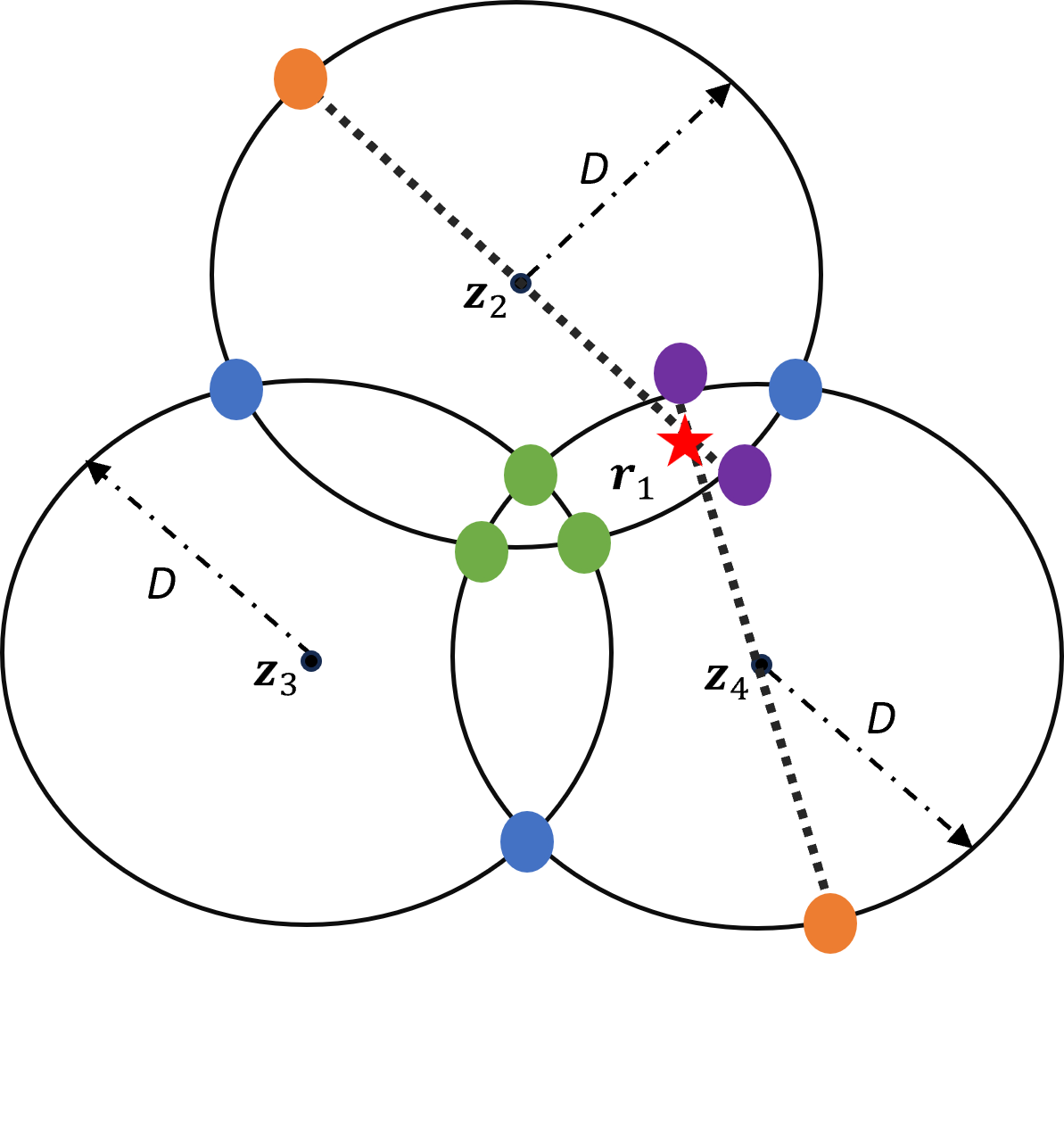}}
        
  	\caption{Examples on the construction of the sets $\mathcal{L}$, $\mathcal{W}_l$, and $\mathcal{U}_l$ with $m=1$ and $M=4$. Blue dots are points in $\mathcal{W}_l$ while green dots are intersections of $\text{Circle}_{l}$ with other circles that do not satisfy $\left\|\mathbf{w}-\mathbf{z}_k\right\|_2 \geq D,~\forall k=1,2, \ldots, M,~k \neq m$. Furthermore, orange dots are elements in $\mathcal{U}_l$ while purple dots are intersections of $\text{Circle}_{l}$ with a line passing through $\mathbf{z}_l$ and $\mathbf{r}_m$ but not satisfying the constraints $\left\|\mathbf{u}-\mathbf{z}_k\right\|_2 \geq D,~\forall k=1,2, \ldots, M,~k \neq m$.} 
    \label{circles}
\end{figure*}
Before solving $\mathcal{P}\text{1-b-m}$, we define the following notations. 
\begin{itemize}
    \item Set $\mathcal{L}$ contains indices $l$ that satisfy $\left\|\mathbf{r}_m-\mathbf{z}_l\right\|_2 < D$.
    \item $\text{Circle}_{l}$ denotes the circle with the center $\mathbf{z}_l$ and radius $D$. 
    \item $\mathcal{W}_l$ denotes the set of points $\mathbf{w}$ that are intersections of $\text{Circle}_{l}$ with $\{\text{Circle}_{k}\}_{k=1, k \neq m,l}^M$ while satisfying the constraints $\left\|\mathbf{w}-\mathbf{z}_k\right\|_2 \geq D,~\forall k=1,2, \ldots, M,~k \neq m,l$. In particular, if the center of $\text{Circle}_{l}$ and that of any circle in $\{ \text{Circle}_{k} \}_{k=1, k \neq m,l}^M$ coincide, the intersection set of these two circles is regarded as empty.
    \item $\mathcal{U}_l$ denotes the set of points $\mathbf{u}$ that are intersections of $\text{Circle}_{l}$ with the line passing through $\mathbf{z}_l$ and $\mathbf{r}_m$ while satisfying the constraints $\left\|\mathbf{u}-\mathbf{z}_k\right\|_2 \geq D,~\forall k=1,2, \ldots, M,~k \neq m,l$. In particular, if $\mathbf{z}_l$ and $\mathbf{r}_m$ coincide, $\mathcal{U}_l$ is regarded as empty.
\end{itemize}

The following two lemmas give a more direct way to construct sets $\mathcal{W}_l$ and $\mathcal{U}_l$.  

\begin{lemma} \label{lemma_w}
(Construction of ${\mathcal{W}}_l$) Given two circles, $\text{Circle}_{l}$ with center $\mathbf{z}_{l} = [x_{l}, y_{l}]^T$ and radius $D$, and $\text{Circle}_{l^{\prime}}$ with center $\mathbf{z}_{l^{\prime}} = [x_{l^{\prime}}, y_{l^{\prime}}]^T$ and radius $D$, the set of intersections of $\text{Circle}_{l}$ and $\text{Circle}_{l^{\prime}}$ is given by
\begin{equation} 
\tilde{\mathcal{W}}_{l,l^{\prime}}=\left\{\begin{array}{l}
\emptyset, ~~~~~~~~~~~\left\|\mathbf{z}_{l} - \mathbf{z}_{l^{\prime}}\right\|_2=0 \text{ or } \left\|\mathbf{z}_{l} - \mathbf{z}_{l^{\prime}}\right\|_2 > 2D, \\
\frac{\mathbf{z}_{l}+\mathbf{z}_{l^{\prime}}}{2} \pm \frac{1}{2} \sqrt{\frac{4 D^2}{\left\|\mathbf{z}_{l} - \mathbf{z}_{l^{\prime}}\right\|_2^2}-1} \begin{bmatrix}
    y_{l^{\prime}}-y_{l} \\
    x_{l}-x_{l^{\prime}}
\end{bmatrix},~\text{otherwise}. \label{Wl}
\end{array}\right. 
\end{equation}
After obtaining the intersections of $\text{Circle}_{l}$ with all other circles (except for $\text{Circle}_{m}$) by~\eqref{Wl}, we obtain $\bigcup_{l^{\prime} \neq m, l}\tilde{\mathcal{W}}_{l,l^{\prime}}$. Then, we remove the points which do not satisfy the constraints $\left\|\mathbf{w}-\mathbf{z}_k\right\|_2 \geq D,~\forall k=1,2, \ldots, M,~k \neq m$ from $\bigcup_{l^{\prime} \neq m, l}\tilde{\mathcal{W}}_{l,l^{\prime}}$ to get ${\mathcal{W}}_l$.
\end{lemma}

\begin{lemma}  \label{lemma_u}
(Construction of ${\mathcal{U}}_l$) Given one circle, $\text{Circle}_{l}$ with center $\mathbf{z}_{l}$ and radius $D$, the set of intersection points of $\text{Circle}_{l}$ with the line passing through $\mathbf{z}_l$ and $\mathbf{r}_m$ is given by
\begin{equation} 
\tilde{\mathcal{U}}_l=\left\{\begin{array}{l}
\emptyset, ~~~~\quad\quad\quad\quad\quad \mathbf{r}_{m} = \mathbf{z}_{l}, \\
\mathbf{z}_l \pm \frac{D(\mathbf{z}_l-\mathbf{r}_m)}{\left\|\mathbf{r}_m-\mathbf{z}_l\right\|_2},~\text{otherwise}. \label{Ul}
\end{array}\right. 
\end{equation}
After obtaining $\tilde{\mathcal{U}}_l$, the points that do not satisfy the distance constraints are removed to obtain ${\mathcal{U}}_l$.
\end{lemma}

To further illustrate the construction of the sets $\mathcal{L}$, $\mathcal{W}_l$, and $\mathcal{U}_l$, Fig.~\ref{circles} shows various scenarios in the case of $m=1$ and $M=4$. For example, in Fig.~\ref{111}, $\mathbf{r}_1$ is more than $D$ away from $\mathbf{z}_2$, $\mathbf{z}_3$ and $\mathbf{z}_4$. Therefore, set $\mathcal{L}$ is empty in this case. On the other hand, for Figs.~\ref{222} and~\ref{333}, $\mathbf{r}_1$ is within distance $D$ from $\mathbf{z}_4$, but not $\mathbf{z}_2$ and $\mathbf{z}_3$, giving $|\mathcal{L}|=1$ in both cases.  Furthermore, for Fig.~\ref{666}, $\mathbf{r}_1$ is within distance $D$ from $\mathbf{z}_2$ and $\mathbf{z}_4$, but not $\mathbf{z}_3$, so $|\mathcal{L}|=2$ in this case.

In Figs.~\ref{222} and~\ref{333}, the green dots and blue dots represent intersections of $\text{Circle}_4$ with other circles, but only the blue dots are points that are at least distance $D$ from other circles' centers. Therefore, only the blue dots are included in the set $\mathcal{W}_4$. Furthermore, the orange and purple dots correspond to the intersections of $\text{Circle}_4$ with the line passing through $\mathbf{z}_4$ and $\mathbf{r}_1$. In Fig.~\ref{222}, the two intersections are included in $\mathcal{U}_4$. While in Fig.~\ref{333}, the purple point needs to be excluded from the candidate set since its distance from the center of $\text{Circle}_2$ is less than $D$.

Finally, in Fig.~\ref{666}, the blue dots on $\text{Circle}_2$ and $\text{Circle}_4$ are elements in $\mathcal{W}_2$ and $\mathcal{W}_4$, respectively. On the other hand, orange dots on $\text{Circle}_2$ and $\text{Circle}_4$ are the elements of $\mathcal{U}_2$ and $\mathcal{U}_4$, respectively.

Then, we discuss the optimal solution of  $\mathbf{z}_m$ in $\mathcal{P}\text{1-b-m}$ by investigating two different cases.

\underline{(a)~$|\mathcal{L}|=0$:} In this case, $\left\|\mathbf{r}_m-\mathbf{z}_l\right\|_2 \geq D$ holds for all $l=1,2, \ldots, M,~l \neq m$. Therefore, all the constraints in $\mathcal{P}\text{1-b-m}$ are satisfied, and the optimal $\mathbf{z}_m^{\star}$ is simply given by $\mathbf{z}_m^{\star} = \mathbf{r}_m$.

\underline{(b)~$|\mathcal{L}| \neq 0$:} In this case, the optimal $\mathbf{z}_m$ is revealed by the following Proposition.

\begin{proposition}
\label{lemma: union of u&w and optimal z}
If $|\mathcal{L}| \neq 0$, the optimal solution of $\mathcal{P}\text{1-b-m}$ is in the set $\bigcup_{l \in \mathcal{L}} \left( \mathcal{W}_l \bigcup \mathcal{U}_l \right)$, and $\mathbf{z}_m^{\star} = \arg \min_{\mathbf{z}_m \in \bigcup_{l \in \mathcal{L}} \left( \mathcal{W}_l \bigcup \mathcal{U}_l \right)} \left\|\mathbf{z}_m - \mathbf{r}_m\right\|_2^2$ is the optimal solution.
\end{proposition}
\begin{proof}
See Appendix~\ref{appendix: optimal solution}.
\end{proof}

The overall procedure for obtaining the optimal solution of $\mathcal{P}\text{1-b-m}$ is summarized in Algorithm~\ref{a2}. For each $\mathcal{P}\text{1-b-m}$, the number of points in the set $\bigcup_{l \in \mathcal{L}} \left( \mathcal{W}_l \bigcup \mathcal{U}_l \right)$ is $c \times 2M$ in the worst case, where $c$ is a constant denoting the number of elements in the set $\mathcal{L}$. Since solving $\mathcal{P}\text{1-b}$ involves solving $M$ subproblems in the form of $\mathcal{P}\text{1-b-m}$, the computational complexity for solving $\mathcal{P}\text{1-b}$ is of order $\mathcal{O}(M^2)$.

\begin{algorithm}[t] 
\caption{Optimal Solution of $\mathcal{P}\text{1-b-m}$ } 
\begin{algorithmic}[1] \label{a2}
\STATE \textbf{Input}: $\mathbf{r}_{m}$, $\{\mathbf{z}_l\}_{l\neq m}$
\STATE Construct the set $\mathcal{L}$.
\STATE \textbf{if} $|\mathcal{L}|=0$
\STATE ~~~~$\mathbf{z}_m^{\star} = \mathbf{r}_m$.
\STATE \textbf{elseif} $|\mathcal{L}|\neq 0$
\STATE ~~~~Construct $\mathcal{W}_l \bigcup \mathcal{U}_l$ for each $l \in \mathcal{L}$.
\STATE ~~~~$\mathbf{z}_m^{\star} = \arg \min_{\mathbf{z}_m \in \bigcup_{l \in \mathcal{L}} \left( \mathcal{W}_l \bigcup \mathcal{U}_l \right)} \left\|\mathbf{z}_m - \mathbf{r}_m\right\|_2^2$.
\STATE \textbf{end}
\STATE \textbf{Output}: $\mathbf{z}_m^{\star}$
\end{algorithmic}
\end{algorithm}


\section{Case Studies} \label{case}
In this section, we apply the proposed optimization framework to solve using three typical problems: capacity maximization, latency minimization, and RZF precoding. Before delving into the details of the case studies, we first introduce the channel model of the MA-aided systems. 


Consider a wireless system with $N$ MAs at the transmitter and $M$ MAs at the receiver. Denote $\mathbf{r}_m = \left[x_{r_m}, y_{r_m} \right]^T$ as the position of the $m$-th MA at the receiver, with $\mathcal{C}_{\text{R}}$ representing the receiver's antenna panel area and the origin is defined at the center of the panel~\cite{MIMO_cap_cha_for_MA}. Similarly, with $\mathcal{C}_{\text{T}}$ denoting the transmitter antenna panel area and the origin defined at the center, $\mathbf{t}_n = \left[x_{t_n}, y_{t_n} \right]^T$ is the position of the $n$-th MA antenna at the transmitter. Using the field response channel model, quasi-static assumption, and far-field assumption~\cite{MIMO_cap_cha_for_MA}, the channel can be formulated as
\begin{align}
   \mathbf{H} \left(  {\{\mathbf{t}_n\}_{n=1}^N,\{\mathbf{r}_m\}_{m=1}^M} \right) = \mathbf{F}^H \mathbf{\Sigma} \mathbf{G},
\label{channel_matrix}
\end{align}
where~$\mathbf{\Sigma} \in \mathbb{C}^{L_r \times L_t}$ is the path response matrix, $L_r$, $L_t$ are the number of receive paths and transmit paths, respectively. Accordingly, the $(q,p)$-th element of $\mathbf{\Sigma}$ denotes the specific response between the $p$-th $\left(p \in \{ 1,2,\dots,L_t \}\right)$ transmit path and the $q$-th $\left(q \in \{ 1,2,\dots,L_r \}\right)$ receive path. Denoting $\theta_{r,q}$ and $\phi_{r,q}$ as the elevation and azimuth angles of the $q$-th receive path, $\mathbf{F} \in \mathbb{C}^{L_r \times M}$, which represents the field response matrix of the receiver, is given by
\begin{align} \label{channel response}
\mathbf{F} &= \exp\left\{j \frac{2 \pi}{\lambda} \begin{bmatrix}
    \boldsymbol{\alpha}_r & \boldsymbol{\beta}_r
\end{bmatrix}  \begin{bmatrix}
    \mathbf{r}_1 & \mathbf{r}_2 & \ldots& \mathbf{r}_M
\end{bmatrix}   \right\} \\
&\triangleq \left[ \mathbf{f}(\mathbf{r}_1),  \mathbf{f}(\mathbf{r}_2), \ldots, \mathbf{f}(\mathbf{r}_M)\right]. \nonumber
\end{align}
In~\eqref{channel response}, $\lambda$ is the carrier wavelength, $\boldsymbol{\alpha}_r = \left[ \sin\theta_{r,1}\cos\phi_{r,1},\right.$ $ \left. \sin\theta_{r,2}\cos\phi_{r,2}, \dots, \sin\theta_{r,L_r}\cos\phi_{r,L_r} \right]^T$ and $\boldsymbol{\beta}_r = \left[ \cos\theta_{r,1},\right.$ $ \left. \cos\theta_{r,2}, \dots, \cos\theta_{r,L_r} \right]^T$. Similarly, letting $\theta_{t,p}$ and $\phi_{t,p}$ denote the departure elevation and azimuth angles of the $p$-th transmit path, the field response matrix of the transmitter $\mathbf{G} \in \mathbb{C}^{L_t \times N}$ is given by
\begin{align}  \label{channel response g}
    \mathbf{G} &= \exp\left\{j \frac{2 \pi}{\lambda} \begin{bmatrix}
    \boldsymbol{\alpha}_t & \boldsymbol{\beta}_t
\end{bmatrix}  \begin{bmatrix}
    \mathbf{t}_1 & \mathbf{t}_2 & \ldots& \mathbf{t}_N
\end{bmatrix}   \right\} \\
&\triangleq \left[ \mathbf{g}(\mathbf{t}_1),  \mathbf{g}(\mathbf{t}_2), \ldots, \mathbf{g}(\mathbf{t}_N)\right], \nonumber
\end{align}
where $\boldsymbol{\alpha}_t = \left[ \sin\theta_{t,1}\cos\phi_{t,1}, \sin\theta_{t,2}\cos\phi_{t,2},\dots,\sin\theta_{t,L_t} \right.$ $\left. \cos\phi_{t,L_t} \right]^T$ and $\boldsymbol{\beta}_t = \left[ \cos\theta_{t,1},\right.$$ \left. \cos\theta_{t,2}, \dots, \cos\theta_{t,L_t} \right]^T$.

\subsection{Capacity Maximization for MA-Aided MIMO System}

This case investigates the channel capacity of a MIMO system with $M$ receive MAs at the BS and a device with $N$ transmit MAs, with both transmit and receiver antenna panels in rectangular shape. Let $\mathbf{s} \in \mathbb{C}^{N}$ denote the transmitted signal from the device. Then, the received signal at the BS is given by
\begin{align}
\mathbf{y}\left({\{\mathbf{t}_n\}_{n=1}^N,\{\mathbf{r}_m\}_{m=1}^M}\right)=\mathbf{H} \left(  {\{\mathbf{t}_n\}_{n=1}^N,\{\mathbf{r}_m\}_{m=1}^M} \right) \mathbf{s}+\boldsymbol{\zeta}, \nonumber
\end{align}
where $\boldsymbol{\zeta} \in \mathbb{C}^{M}$ contains independent and identically distributed (i.i.d.) Gaussian noise elements, each following $\mathcal{CN} \left(0, \sigma^{2}\right)$ with $\sigma^{2}$ being the noise power.  Assuming that perfect channel state information is available at both the device and BS, the problem of capacity maximization is formulated as
\begin{subequations}
\begin{align}
&\mathcal{P}\text {(A):} \min_{\{\mathbf{t}_n\}_{n=1}^N,\{\mathbf{r}_m\}_{m=1}^M, \mathbf{Q}}~  -\log _2 \operatorname{det}\left(\mathbf{I}_M+\frac{1}{\sigma^2} \mathbf{H} \mathbf{Q H}^H\right) \\
&\text { s.t. }~~~~ 
\operatorname{Tr}(\mathbf{Q}) \leq P_{\text{max}}, \label{tr_Q} \\
&~~~~~~~~~~\mathbf{Q} \succeq \mathbf{0}, \label{Q_positive} \\
&~~~~~~~~~~~
\mathbf{r}_m \in \mathcal{C}_{\text{R}}, ~\forall m =1,2, \ldots, M, \label{given_region_r} \\
&~~~~~~~~~~
\left\|\mathbf{r}_m-\mathbf{r}_l\right\|_2 \geq D,~\forall m,l=1,2, \ldots, M, ~m \neq l, \label{antenna_distance_r} \\
&~~~~~~~~~~~
\mathbf{t}_n \in \mathcal{C}_{\text{T}}, ~\forall n =1,2, \ldots, N, \label{given_region_t} \\
&~~~~~~~~~~
\left\|\mathbf{t}_n-\mathbf{t}_l\right\|_2 \geq D,~\forall n,l=1,2, \ldots, N, ~n \neq l, \label{antenna_distance_t}
\end{align}
\end{subequations}
where $\mathbf{Q} \triangleq \mathbb{E}\left\{\mathbf{s} \mathbf{s}^H\right\} $ denotes the transmit covariance matrix and $P_{\text{max}}$ is the maximum transmit power of the device. For notational simplicity, we use $\mathbf{H}$ to represent $\mathbf{H} \left(  {\{\mathbf{t}_n\}_{n=1}^N,\{\mathbf{r}_m\}_{m=1}^M} \right)$. Obviously, $\mathcal{P}\text {(A)}$ is in the same form as the general problem formulation $\mathcal{P}$, with $\mathbf{X} = \mathbf{Q}$, set $\mathcal{X}$ determined by~\eqref{tr_Q},~\eqref{Q_positive}, and the MA positions are $\{\mathbf{r}_m\}_{m=1}^M$, $\{\mathbf{t}_n\}_{n=1}^N$.

According to the proposed penalty framework in Section~\ref{framework}, by introducing auxiliary variables $\{\mathbf{z}_m = \mathbf{r}_m\}_{m=1}^M$ and $\{\mathbf{u}_n = \mathbf{t}_n\}_{n=1}^N$, $\mathcal{P}\text {(A)}$ becomes $\mathcal{P}\text {(A1)}$  on top of the next page.
\begin{figure*}
\begin{align}
\mathcal{P}\text {(A1)}: &\min_{\{\mathbf{r}_m, \mathbf{z}_m\}_{m=1}^M, \{\mathbf{t}_n, \mathbf{u}_n\}_{n=1}^N, \mathbf{Q}}  -\log _2 \operatorname{det}\left(\mathbf{I}_M+\frac{1}{\sigma^2} \mathbf{H} \mathbf{Q H}^H\right)+\rho \left( \sum_{m=1}^M \left\|\mathbf{r}_m - \mathbf{z}_m\right\|_2^2 + \sum_{n=1}^N \left\|\mathbf{t}_n - \mathbf{u}_n\right\|_2^2 \right) \label{ori_a11}\\
&~~~~~~~~~~~~\text { s.t. } 
~~~~~~ \eqref{tr_Q}, ~\eqref{Q_positive}, ~\eqref{given_region_r}, \text{ and } \eqref{given_region_t}, \nonumber \\
&\quad~~~~~~~~~~~~~~~~~~~~ \left\|\mathbf{z}_m-\mathbf{z}_l\right\|_2 \geq D,  ~~~~\forall m,l=1,2, \ldots, M, ~~ m \neq l, \nonumber \\
&\quad~~~~~~~~~~~~~~~~~~~~ \left\|\mathbf{u}_n-\mathbf{u}_l\right\|_2 \geq D,  ~~~~\forall n,l=1,2, \ldots, N, ~~ n \neq l. \nonumber
\end{align} 
\hrulefill
\begin{align}
\nabla_{\mathbf{r}_m} f_{\mathcal{P}\text {(A1)}} &= \frac{-4 \pi \lambda^{-1} \sigma^{-2} \operatorname{det}\left(\mathbf{I}_N +\frac{1}{\sigma^2}\mathbf{W}_{\backslash m}^H\mathbf{W}_{\backslash m}\right)}{\ln 2 \cdot \operatorname{det} \left(\mathbf{I}_M+\frac{1}{\sigma^2} \mathbf{H} \mathbf{Q H}^H\right)} \Re \left\{ 
\begin{bmatrix}
    \boldsymbol{\alpha}_r & \boldsymbol{\beta}_r
\end{bmatrix}^T \left[ \left(j\mathbf{B}_{\backslash m} \mathbf{f}(\mathbf{r}_m) \right) \circ \mathbf{f}(\mathbf{r}_m)^* \right] \right\} + 2\rho \left( \mathbf{r}_m - \mathbf{z}_m \right) \label{g_r} \\ 
\nabla_{\mathbf{t}_n} f_{\mathcal{P}\text {(A1)}} &= \frac{-4 \pi \lambda^{-1} \sigma^{-2} \operatorname{det}\left(\mathbf{I}_M +\frac{1}{\sigma^2}\mathbf{P}_{\backslash n}^H\mathbf{P}_{\backslash n}\right)}{\ln 2 \cdot \operatorname{det} \left(\mathbf{I}_M+\frac{1}{\sigma^2} \mathbf{H} \mathbf{Q H}^H\right)} \Re \left\{ \begin{bmatrix}
    \boldsymbol{\alpha}_t & \boldsymbol{\beta}_t
\end{bmatrix}^T  \left[ \left(j\mathbf{D}_{\backslash n} \mathbf{g}(\mathbf{t}_n) \right) \circ \mathbf{g}(\mathbf{t}_n)^* \right] \right\} + 2\rho \left( \mathbf{t}_n - \mathbf{u}_n \right) \label{g_t}
\end{align}
\hrulefill
\end{figure*}
The problem $\mathcal{P}\text {(A1)}$ is then solved by alternately optimizing the blocks $\mathbf{Q}$, $\{\mathbf{t}_n\}_{n=1}^N$, $\{\mathbf{r}_m\}_{m=1}^M$, $\{\mathbf{z}_m\}_{m=1}^M$, and $\{\mathbf{u}_n\}_{n=1}^N$. Since the optimization of $\{\mathbf{u}_n\}_{n=1}^N$ and $\{\mathbf{z}_m\}_{m=1}^M$ can be solved using Algorithm~\ref{a2} in Section~\ref{framework}, we only present the details about the optimization of $\mathbf{Q}$, $\{\mathbf{r}_m\}_{m=1}^M$, and $\{\mathbf{t}_n\}_{n=1}^N$ here.
 
\underline{(a)~Subproblem with respect to $\mathbf{Q}$~\cite{MIMO_cap_cha_for_MA}:} 
With $\{\mathbf{t}_n,\mathbf{u}_n\}_{n=1}^N$ and $\{\mathbf{r}_m,\mathbf{z}_m\}_{m=1}^M$ fixed, the optimization of $\mathbf{Q}$ has no difference from traditional resource allocation problem. In particular, by denoting $\mathbf{H}=\tilde{\mathbf{U}} \tilde{\boldsymbol{\Lambda}} \tilde{\mathbf{V}}^H$ as the truncated singular
value decomposition of $\mathbf{H}$, where $\tilde{\mathbf{U}} \in \mathbb{C}^{M \times S}$, $\tilde{\boldsymbol{\Lambda}} \in \mathbb{C}^{S \times S}$, $\tilde{\mathbf{V}} \in \mathbb{C}^{N \times S}$, and $S = \text{rank}\left(\mathbf{H}\right)$, the optimal $\mathbf{Q}^{\star}$ is given by
\begin{align}  \label{optimal_Q}
\mathbf{Q}^{\star}=\tilde{\mathbf{V}} \operatorname{diag}\left(\left[\gamma_1^{\star}, \gamma_2^{\star}, \ldots, \gamma_S^{\star}\right]\right) \tilde{\mathbf{V}}^H,
\end{align}
where $\gamma_s^{\star}=\max \left(1/ \gamma_0-\sigma^2 / \tilde{\boldsymbol{\Lambda}}[s, s]^2, 0\right)$ with $\sum_{s=1}^S \gamma_s^{\star}= P_{\text{max}}$, and $\gamma_0$ is a constant so-called `water-filling level'~\cite{water-filling}.

\underline{(b)~Subproblem with respect to $\{\mathbf{r}_m\}_{m=1}^M$:} 
Since the objective function~\eqref{ori_a11} is differentiable with respect to $\{\mathbf{r}_m\}_{m=1}^M$, we apply the projected gradient approach to optimize $\{\mathbf{r}_m\}_{m=1}^M$. For each $m$ and at iteration $i$,
\begin{align} \label{pgm_r}
\mathbf{r}_m^{(i)} = \mathcal{P}_{\mathcal{C}_{\text{R}}}\left\{\mathbf{r}_m^{(i-1)} - \eta^{(i-1)} \nabla_{\mathbf{r}_m} f_{\mathcal{P}\text {(A1)}} \right\},
\end{align}
where $f_{\mathcal{P}\text {(A1)}}$ denotes the objective function of $\mathcal{P}\text {(A1)}$, and the projection is given by $\mathcal{P}_{\mathcal{C}_{\text{R}}} \left( \mathbf{x} \right) \triangleq \operatorname{min} \left( \operatorname{max} \left( \mathbf{x}, \mathbf{x}_{\text{min}} \right), \mathbf{x}_{\text{max}} \right)$ with $\mathbf{x}_{\text{min}}$ and $\mathbf{x}_{\text{max}}$ denoting the lower left and upper right coordinates of $\mathcal{C}_{\text{R}}$. Additionally, the gradient $\nabla_{\mathbf{r}_m} f_{\mathcal{P}\text {(A1)}}=[\frac{\partial f_{\mathcal{P}\text {(A1)}}}{\partial x_{r_m}}, \frac{\partial f_{\mathcal{P}\text {(A1)}}}{\partial y_{r_m}}]^T$ is computed by chain rules, and the result is given by \eqref{g_r} with $\mathbf{B}_{\backslash m} = \mathbf{\Sigma G}\mathbf{U}_Q \mathbf{V}_Q^{\frac{1}{2}} $ $ \left(\mathbf{I}_N +\frac{1}{\sigma^2} \mathbf{W}_{\backslash m}^H \mathbf{W}_{\backslash m} \right)^{-1} \mathbf{V}_Q^{\frac{1}{2}} \mathbf{U}_Q^H \mathbf{G}^H \mathbf{\Sigma}^H$, where $\mathbf{V}_Q$ and $\mathbf{U}_Q$ are obtained by the eigenvalue decomposition $\mathbf{Q} = \mathbf{U}_Q \mathbf{V}_Q \mathbf{U}_Q^H$. On the other hand, $\mathbf{W}_{\backslash m}$ is $\mathbf{F}^H \mathbf{\Sigma} \mathbf{G} \mathbf{U}_Q \mathbf{V}_Q^{\frac{1}{2}}$ but with its $m$-th row removed.

\underline{(c)~Subproblem with respect to $\{\mathbf{t}_n\}_{n=1}^N$:}  Since this subproblem has the same structure as for $\{\mathbf{r}_m\}_{m=1}^M$, for each $n$ and iteration $i$, $\mathbf{t}_n^{(i)}$ is updated by
\begin{align} \label{pgm_t}
\mathbf{t}_n^{(i)} = \mathcal{P}_{\mathcal{C}_{\text{T}}}\left\{\mathbf{t}_n^{(i-1)} - \eta^{(i-1)} \nabla_{\mathbf{t}_n} f_{\mathcal{P}\text {(A1)}} \right\},
\end{align}
where the projection $\mathcal{P}_{\mathcal{C}_{\text{T}}}$ is similar to $\mathcal{P}_{\mathcal{C}_{ \text{R}}}$ defined above, and the gradient $\nabla_{\mathbf{t}_n} f_{\mathcal{P}\text {(A1)}}=[\frac{\partial f_{\mathcal{P}\text {(A1)}}}{\partial x_{t_n}}, \frac{\partial f_{\mathcal{P}\text {(A1)}}}{\partial y_{t_n}}]^T$ is given by \eqref{g_t}. Note that $\mathbf{D}_{\backslash n} = \mathbf{\Sigma}^H \mathbf{F} \mathbf{U}_S \mathbf{V}_S^{\frac{1}{2}}\left(\mathbf{I}_M +\frac{1}{\sigma^2}\mathbf{P}_{\backslash n}^H\mathbf{P}_{\backslash n} \right)^{-1} $ $\mathbf{V}_S^{\frac{1}{2}} \mathbf{U}_S^H \mathbf{F}^H \mathbf{\Sigma}$, where $\mathbf{V}_S$ and $\mathbf{U}_S$ are obtained by the eigenvalue decomposition of the equivalent covariance matrix, defined in the Section III of~\cite{MIMO_cap_cha_for_MA}. $\mathbf{P}_{\backslash n}$ is $\mathbf{G}^H \mathbf{\Sigma}^H \mathbf{F} \mathbf{U}_S \mathbf{V}_S^{\frac{1}{2}}$ but with its $n$-th row removed.

In summary, the procedure for solving $\mathcal{P}\text {(A)}$ is given in Algorithm~\ref{algorithm:case1}.
\begin{algorithm}[t] 
\caption{The Penalty Method for Solving $\mathcal{P}\text {(A)}$ } 
\begin{algorithmic}[1] \label{algorithm:case1}
\STATE Initialize $\mathbf{Q}$, $\{\mathbf{t}_n\}_{n=1}^N$, $\{\mathbf{r}_m\}_{m=1}^M$, $\{\mathbf{z}_m\}_{m=1}^M$, and $\{\mathbf{u}_n\}_{n=1}^N$.
\REPEAT
\STATE Optimize $\mathbf{Q}$ according to~\eqref{optimal_Q}.
\STATE \textbf{for} $m=1,\ldots,M$ \textbf{do}
\STATE ~~Optimize $\{\mathbf{r}_m\}_{m=1}^M$ with iteration on~\eqref{pgm_r} with respect to $i$.
\STATE \textbf{end for}
\STATE \textbf{for} $n=1,\ldots,N$ \textbf{do}
\STATE ~~Optimize $\{\mathbf{t}_n\}_{n=1}^N$ with iteration on~\eqref{pgm_t} with respect to $i$.
\STATE \textbf{end for}
\STATE \textbf{for} $m=1,\ldots,M$ \textbf{do}
\STATE ~~~~Optimize $\mathbf{z}_m$ using Algorithm~\ref{a2}.
\STATE \textbf{end for}
\STATE \textbf{for} $n=1,\ldots,N$ \textbf{do}
\STATE ~~~~Optimize $\mathbf{u}_n$ using Algorithm~\ref{a2} with input $\mathbf{r}_m$ replaced by $\mathbf{t}_n$ and $\mathbf{z}_l$ replaced by $\mathbf{u}_l$.
\STATE \textbf{end for}
\STATE Increase the penalty factor $\rho$.
\UNTIL Stopping criterion is satisfied.	
\end{algorithmic}
\end{algorithm}

\subsection{Latency Minimization for MA-Aided Mobile Edge Computing (MEC)}

This case study focuses on the latency minimization for a single-server MEC network. The considered mobile network consists of $N$ single-FPA users and a BS with $M$ MAs. In this case, $\{\mathbf{t}_n\}_{n=1}^N$ in~\eqref{channel_matrix} are the locations of the users assumed to be known and thus do not get optimized. With $\mathbf{s} \in \mathbb{C}^{N}$ being the transmitted signals from all users each with unit power, the received signal at the BS is expressed as
\begin{align}
    \mathbf{y}\left(\{\mathbf{r}_m\}_{m=1}^M\right)=\mathbf{W}^H\mathbf{H}\left(\{\mathbf{r}_m\}_{m=1}^M\right)\mathbf{P} ^{\frac{1}{2}}\mathbf{s}+\mathbf{W}^H\boldsymbol{\zeta},
\end{align}
where $\mathbf{P} \in \mathbb{R}^{N \times N}$ is a diagonal matrix whose elements, $\{p_n\}_{n=1}^N$, are the transmit power of users, $\mathbf{H}\left(\{ \mathbf{r}_m \}_{m=1}^M\right)$ is given by~\eqref{channel_matrix} but with $\{\mathbf{t}_n\}_{n=1}^N$ fixed, and $\mathbf{W}\in \mathbb{C}^{M \times N}$ denotes the receive beamforming matrix at the BS. 

We assume all users are executing their tasks separately, and these tasks can either be computed locally with the results subsequently sent to the server by the users or be offloaded to the MEC server for computing. The offloading scheme is binary offloading controlled by the offloading parameter $\beta_n$ for user $n$ with $\beta_n=0$ representing that user $n$ executes the task locally and $\beta_n=1$ representing that user $n$ offloads the task to the MEC server.

If user $n$ chooses to compute the task locally, the latency experienced by it is given by the sum of the local computing latency and the uploading latency, i.e.,
\begin{align}
    T_{n,\text{local}} = \frac{D_n C_n}{f_n^{\text{loc}}} + \frac{V_n}{R_n},
\end{align}
where $D_n$ is the data size of the task of user $n$, $C_n$ refers to the number of central processing unit (CPU) cycles required to locally process one bit of data for user $n$, $f_n^{\text{loc}}$ denotes the CPU frequency of user $n$, and $V_n$ represents the amount of data for the computation result, which is assumed to be proportional to $D_n$. Moreover, the transmission rate $R_n$ is given by
\begin{align}
R_n = b \log_2 \left(1+\frac{|\mathbf{w}_n^H \mathbf{h}_n(\{\mathbf{r}_m\}_{m=1}^M)|^2 p_n}{\sum_{i\neq n} |\mathbf{w}_n^H \mathbf{h}_i(\{\mathbf{r}_m\}_{m=1}^M)|^2 p_i + ||\mathbf{w}_n||_2^2\sigma^2}\right), \nonumber
\end{align}
where $b$ refers to the bandwidth, $\mathbf{h}_n(\{\mathbf{r}_m\}_{m=1}^M)$ and $\mathbf{w}_n$ are the $n$-th column of $\mathbf{H}\left(\{ \mathbf{r}_m\}_{m=1}^M \right)$ and $\mathbf{W}$, respectively.

On the other hand, if user $n$ chooses to offload the computation task, the latency is the sum of the offloading latency and the edge computing latency, which is expressed as
\begin{align}
    T_{n,\text{edge}} = \frac{D_n}{R_n} + \frac{D_n C_s}{f_n^\text{s}}, \nonumber
\end{align}
where $C_s$ represents the number of CPU cycles required to process one bit of data at the MEC server, and $f_n^\text{s}$ is the CPU frequency of the server assigned to user $n$. 

In binary offloading setting, the total latency of network is expressed as 
\begin{align}
    T(\{\mathbf{r}_m\}_{m=1}^M,\boldsymbol{\beta},\mathbf{f}^\text{s}) = \sum_{n=1}^N \left[ \left(1-\beta_n \right)T_{n,\text{local}} + \beta_n T_{n,\text{edge}} \right], \nonumber
\end{align}
where $\boldsymbol{\beta}=\left[\beta_1, \beta_2, \ldots, \beta_N\right]^T$ and $\mathbf{f^\text{s}}=\left[f^\text{s}_1, f^\text{s}_2, \ldots, f^\text{s}_N\right]^T$. To be consistent with~\cite{FA_MEC}, $\mathbf{W}$ is assumed to be the zero-forcing (ZF) combing~\cite[eq.~(4)]{FA_MEC}. Thus, the problem of latency minimization can be formulated as
\begin{subequations}
\begin{align}
&\mathcal{P}\text {(B):}~~~\min_{\{\mathbf{r}_m\}_{m=1}^M,\boldsymbol{\beta},\mathbf{f^\text{s}}} \quad T(\{\mathbf{r}_m\}_{m=1}^M,\boldsymbol{\beta},\mathbf{f}^s) \label{latency_o.f.} \\ 
&\text { s.t. }~~~
\beta_n \in  \left\{0,1  \right\}, ~~n=1,2,\dots,N, \label{binary_constraint}  \\
& ~~~~~~~~ \sum_{n=1}^N f_n^\text{s} \leq f_{\text{server}}, \label{constraint_fs} \\
&~~~~~~~~~
\mathbf{r}_m \in \mathcal{C}_{\text{R}}, ~\forall m =1,2, \ldots, M, \nonumber \\
&~~~~~~~~
\left\|\mathbf{r}_m-\mathbf{r}_l\right\|_2 \geq D,~\forall m,l=1,2, \ldots, M, ~m \neq l,  \nonumber
\end{align}
\end{subequations}
where \eqref{binary_constraint} imposes a binary constraint on the offloading option for each user, and the second constraint ensures that the allocated computational resources of the MEC server do not exceed its maximal frequency $f_{\text{server}}$. This optimization problem is evidently non-convex and NP-hard. One existing method \cite{FA_MEC} splits all variables into two blocks: ${\boldsymbol{\beta}, \mathbf{f^\text{s}}}$ and $\{\mathbf{r}_m\}_{m=1}^M$. However, given fixed $\boldsymbol{\beta}$ and $\mathbf{f^\text{s}}$, the subproblem with respect to $\{\mathbf{r}_m\}_{m=1}^M$ faces the constraint~\eqref{given_region} and~\eqref{antenna_distance}. To this end,~\cite{FA_MEC} tried to overcome the optimization challenge by using the particle swarm optimization (PSO) algorithm. As $\mathcal{P}\text {(B)}$ is in the form of the general optimization problem $\mathcal{P}$, we demonstrate how the proposed framework can handle this MA-aided latency minimization problem.

By employing the proposed framework in Section~\ref{framework}, problem $\mathcal{P}\text {(B)}$ can be transformed into $\mathcal{P}\text{(B1)}$ on top of the next page.
\begin{figure*}
\begin{align}
\mathcal{P}\text {(B1):} &\min_{\{\mathbf{r}_m, \mathbf{z}_m\}_{m=1}^M,\boldsymbol{\beta},\mathbf{f^\text{s}}} \quad T(\{\mathbf{r}_m\}_{m=1}^M,\boldsymbol{\beta},\mathbf{f}^\text{s}) + \rho \sum_{m=1}^M \left\|\mathbf{r}_m - \mathbf{z}_m\right\|_2^2 \label{of_PC2} \\
&~~~~~~~\text { s.t. } ~~~~~~~~~ \eqref{binary_constraint}, ~\eqref{constraint_fs}, \nonumber \\
&\quad~~~~~~~~~~~~~~~~~~~
\mathbf{r}_m \in \mathcal{C}_{\text{R}}, ~\forall m =1,2, \ldots, M, \nonumber \\
&\quad~~~~~~~~~~~~~~~~~~ \left\|\mathbf{z}_m-\mathbf{z}_l\right\|_2 \geq D,  ~~~~\forall m,l=1,2, \ldots, M, ~~ m \neq l.  \nonumber
\end{align} 
\hrulefill
\begin{align}
\nabla_{\mathbf{r}_m} f_{\mathcal{P}\text {(B1)}} &= \sum_{n=1}^N  \frac{-4 \pi p_n \lambda^{-1} \sigma^{-2} c_n \ln 2}{k_n\left( \ln k_n\right)^2} \Re \left\{ 
\begin{bmatrix}
    \boldsymbol{\alpha}_r & \boldsymbol{\beta}_r
\end{bmatrix}^T \left[ \left(j \mathbf{\Sigma} \mathbf{g}\left(\mathbf{t}_n \right) \mathbf{g}\left(\mathbf{t}_n \right)^H \mathbf{\Sigma}^H \mathbf{f}(\mathbf{r}_m) \right) \circ \mathbf{f}(\mathbf{r}_m)^* \right] \right\} + 2\rho \left( \mathbf{r}_m - \mathbf{z}_m \right) \label{g_r-2}
\end{align}
\hrulefill
\end{figure*}
$\mathcal{P}\text {(B1)}$ can also be solved by alternately optimizing the blocks $\boldsymbol{\beta}$, $\mathbf{f^\text{s}}$, $\{\mathbf{r}_m\}_{m=1}^M$, and $\{\mathbf{z}_m\}_{m=1}^M$. As the optimization with respect to $\{\mathbf{z}_m\}_{m=1}^M$ can be solved by Algorithm~\ref{a2}, we only present the details about the optimization of $\boldsymbol{\beta}$, $\mathbf{f^\text{s}}$, and $\{\mathbf{r}_m\}_{m=1}^M$.

\underline{(a)~Subproblem with respect to $\boldsymbol{\beta}$:}
In general, the binary constraint \eqref{binary_constraint} can be handled by continuous relaxation followed by projection~\cite{FA_MEC} or the penalty method~\cite{zhenrong}. However, we note that the subproblem after relaxation with respect to $\boldsymbol{\beta}$ is a linear program with box constraint $0 \leq \beta_n \leq 1,~~n=1,2,\dots,N$. Leveraging this observation, the subproblem here has a closed-form solution: 
\begin{align}   \label{optimal_beta}
    \beta_n^{\star} = \left\{ 
    \begin{aligned}
    &0, \quad  (D_n - V_n) f_n^\text{s} f_n^\text{loc} + D_n C_s R_n f_n^\text{loc} \geq D_n C_n R_n f_n^\text{s} \\
    &1, \quad  (D_n - V_n) f_n^\text{s} f_n^\text{loc} + D_n C_s R_n f_n^\text{loc} < D_n C_n R_n f_n^\text{s}.
    \end{aligned}
    \right. 
\end{align}

\underline{(b)~Subproblem with respect to $\mathbf{f^\text{s}}$:} 
According to the method proposed in \cite{joint_TOandRA_for_MSMEC}, this subproblem has closed-form solution:
\begin{align} \label{optimal_fs}
    f^{\text{s}*} = \frac{f_{\text{server}}\sqrt{f_n^{\text{loc}}}}{\sum_{n \in \mathcal{N}_\text{s}} \sqrt{f_n^{\text{loc}}} },
\end{align}
where the set $\mathcal{N}_\text{s}$ refers to the collection of users who opt to offload their tasks to the MEC server.

\underline{(c)~Subproblem with respect to $\{\mathbf{r}_m\}_{m=1}^M$:}
Since the objective function~\eqref{of_PC2} is differentiable with respect to $\{\mathbf{r}_m\}_{m=1}^M$, we apply the projected gradient approach to optimize $\{\mathbf{r}_m\}_{m=1}^M$, i.e.,
\begin{align} \label{pgm_r_case2}
\mathbf{r}_m^{(i)} = \mathcal{P}_{\mathcal{C}_{\text{R}}}\left\{\mathbf{r}_m^{(i-1)} - \eta^{(i-1)} \nabla_{\mathbf{r}_m} f_{\mathcal{P}\text {(B1)}} \right\},
\end{align}
where the gradient can be obtained by chain rules, which is given by \eqref{g_r-2}, with $c_n = \left(1-\beta_n \right)V_n + \beta_n D_n$ and $k_n = 1+\frac{p_n}{\sigma^2} \mathbf{g}\left(\mathbf{t}_n \right)^H \mathbf{\Sigma}^H \mathbf{F}\left(\{\mathbf{r}_m\}_{m=1}^M\right) \mathbf{F}\left(\{\mathbf{r}_m\}_{m=1}^M\right)^H \mathbf{\Sigma} \mathbf{g}\left(\mathbf{t}_n \right)$.



The overall algorithm for solving $\mathcal{P}\text {(B)}$ is summarized in Algorithm~\ref{algorithm:case2}.
\begin{algorithm}[t] 
\caption{The Penalty Method for Solving $\mathcal{P}\text {(B)}$ } 
\begin{algorithmic}[1] \label{algorithm:case2}
\STATE Initialize the $\{\mathbf{r}_m\}_{m=1}^M$, $\{\mathbf{z}_m\}_{m=1}^M$, $\boldsymbol{\beta}$, and $\mathbf{f}^{\text{s}}$.
\REPEAT
\STATE Optimize $\boldsymbol{\beta}$ according to~\eqref{optimal_beta}.
\STATE Optimize $\mathbf{f}^{\text{s}}$ according to~\eqref{optimal_fs}.
\STATE \textbf{for} $m=1,\ldots,M$ \textbf{do}
\STATE ~~Optimize $\{\mathbf{r}_m\}_{m=1}^M$ with iteration on~\eqref{pgm_r_case2} with respect to $i$.
\STATE \textbf{end for}
\STATE \textbf{for} $m=1,\ldots,M$ \textbf{do}
\STATE ~~~~Optimize $\mathbf{z}_m$ using Algorithm~\ref{a2}.
\STATE \textbf{end for}
\STATE Increase the penalty factor $\rho$.
\UNTIL Stopping criterion is satisfied.	
\end{algorithmic}
\end{algorithm}

\subsection{Regularized Zero-Forcing Precoding for MA-Aided Multi-User System}

This case considers a multi-user multiple-input single-output (MISO) downlink communication system with $N$ transmit MAs at the BS and $M$ single-FPA users. In this case, the channel matrix $\mathbf{H}$ only depends on $\{ \mathbf{t}_n \}_{n=1}^N$, while $\{ \mathbf{r}_m \}_{m=1}^M$ are the locations of the users which are assumed to be known. The received signal at the $m$-th user is then given by 
\begin{align}
y_m = \mathbf{h}_m\left(\{\mathbf{t}_n\}_{n=1}^N\right)^H \mathbf{W} \mathbf{s}+ \zeta_m,
\end{align}
where $\mathbf{s} \in \mathbb{C}^{M}$ represents the transmitted data for all $M$ users with $ \mathbb{E}\left\{\mathbf{s} \mathbf{s}^H\right\}=\mathbf{I}_M$, $\mathbf{W} \triangleq [\mathbf{w}_1,\mathbf{w}_2,...,\mathbf{w}_N]^H \in \mathbb{C}^{N \times M}$ is the precoding matrix, $\mathbf{h}_m\left(\{\mathbf{t}_n\}_{n=1}^N\right)^H$ denotes the m-th row of the channel matrix $\mathbf{H}\left(\{\mathbf{t}_n\}_{n=1}^N\right)$ and $\zeta_m \sim \mathcal{C}\mathcal{N}\left(0, \sigma^{2}_m\right)$ denotes the Gaussian noise. 


The RZF problem is thus formulated as
\begin{align}
&\mathcal{P}\text {(C):}~~~\min_{\{\mathbf{t}_n\}_{n=1}^N, \mathbf{W}} \quad \left\|\mathbf{I}_M-\mathbf{H} \left(\{\mathbf{t}_n\}_{n=1}^N\right)\mathbf{W}\right\|_F^2+\alpha\|\mathbf{W}\|_F^2 \nonumber \\
&\text { s.t. }~~~~~~
\mathbf{t}_n \in \mathcal{C}_{\text{T}}, ~\forall n =1,2, \ldots, N, \nonumber \\
&~~~~~~~~~~\left\|\mathbf{t}_n - \mathbf{t}_l\right\|_2 \geq D,~\forall n,l=1,2, \ldots, N, ~n \neq l, \nonumber
\end{align}
where $\alpha$ is a hyperparameter controlling the flexibility of precoding. An existing approach~\cite{Flexible_precoding_MU_MAcomm} transforms $\mathcal{P}\text {(C)}$ to a sparse optimization problem. However, this transformation introduces additional zero-norm constraint, which is challenging to tackle and exacerbates the non-convexity of the transformed problem. To deal with it ,~\cite{Flexible_precoding_MU_MAcomm} subsequently leverages the compressed sensing (CS) approach, specifically, the offgrid regularized least squares-based orthogonal matching pursuit (RLS-OMP) algorithm to solve the problem. Next, we demonstrate how the proposed framework facilitates in solving problem $\mathcal{P}\text {(C)}$.

By applying the proposed framework, problem $\mathcal{P}\text {(C)}$ is converted into $\mathcal{P}\text {(C1)}$ on top of the next page.
\begin{figure*}
\begin{align}
\mathcal{P}\text {(C1):}~~~&\min_{\{\mathbf{t}_n, \mathbf{u}_n\}_{n=1}^N, \mathbf{W}} \quad \left\|\mathbf{I}_M-\mathbf{H} \left(\{\mathbf{t}_n\}_{n=1}^N\right)\mathbf{W}\right\|_F^2+\alpha\|\mathbf{W}\|_F^2 +\rho \sum_{n=1}^N \left\|\mathbf{t}_n - \mathbf{u}_n\right\|_2^2 \label{o1}\\ 
&~~~~~\text { s.t. } ~~~~~~
\mathbf{t}_n \in \mathcal{C}_{\text{T}}, ~\forall n =1,2, \ldots, N, \nonumber \\
&\quad~~~~~~~~~~~~~ \left\|\mathbf{u}_n-\mathbf{u}_l\right\|_2 \geq D,  ~~~~\forall n,l=1,2, \ldots, N, ~~ n \neq l. \nonumber
\end{align} 
\hrulefill
\begin{align}
\nabla_{\mathbf{t}_n} f_{\mathcal{P}\text {(C1)}} &= \frac{4 \pi}{\lambda} \Re \left\{ \begin{bmatrix}
    \boldsymbol{\alpha}_t & \boldsymbol{\beta}_t
\end{bmatrix}^T  \left[j \left(- \mathbf{w}_n^H \otimes \mathbf{B}^H \operatorname{vec}\left(\mathbf{A}\right) + \mathbf{w}_n^H \otimes \left( \mathbf{B}^H\mathbf{B}\mathbf{g}(\mathbf{t}_n) \right) \mathbf{w}_n \right) \circ \mathbf{g}(\mathbf{t}_n)^* \right] \right\} + 2\rho \left( \mathbf{t}_n - \mathbf{u}_n \right). \label{g_t-case3}
\end{align}
\hrulefill
\end{figure*}
Similar to the previous two case studies, problem $\mathcal{P}\text {(C1)}$ can be solved by alternately optimizing the blocks $\mathbf{W}$, $\{\mathbf{t}_n\}_{n=1}^N$, and $\{\mathbf{u}_n\}_{n=1}^N$. The details of optimization are provided below.

\underline{(a)~Subproblem with respect to $\mathbf{W}$:}
According to the~\cite[eq.~(5)]{Flexible_precoding_MU_MAcomm}, when fixing $\{\mathbf{t}_n\}_{n=1}^N$ and $\{\mathbf{u}_n\}_{n=1}^N$, the optimization for $\mathbf{W}$ has a closed-form solution given by 
\begin{align} \label{optimal_W}
    \mathbf{W}^{\star} = \mathbf{H}\left(\{\mathbf{t}_n\}_{n=1}^N\right)^H\left(\mathbf{H}\left(\{\mathbf{t}_n\}_{n=1}^N\right)\mathbf{H}\left(\{\mathbf{t}_n\}_{n=1}^N \right)^H + \alpha \mathbf{I}\right)^{-1}. 
\end{align}

\underline{(b)~Subproblem with respect to $\{\mathbf{t}_n\}_{n=1}^N$:}
The projected gradient approach can be employed to optimize $\{\mathbf{t}_n\}_{n=1}^N$ because the objective function~\eqref{o1} is differentiable with respect to $\{\mathbf{t}_n\}_{n=1}^N$, i.e.,
\begin{align} \label{pgm_t_case3}
\mathbf{t}_n^{(i)} = \mathcal{P}_{\mathcal{C}_{\text{T}}}\left\{\mathbf{t}_n^{(i-1)} - \eta^{(i-1)} \nabla_{\mathbf{t}_n} f_{\mathcal{P}\text {(C1)}} \right\},
\end{align}
where the gradient is given by \eqref{g_t-case3}. Note that $\mathbf{A} = \mathbf{I}_M - \Sigma_{k=1,k\neq n}^N \mathbf{B} \mathbf{g}\left( \mathbf{t}_k \right) \mathbf{w}_k^H$, where $\mathbf{B} = \mathbf{F}^H\mathbf{\Sigma}$. In particular, $\otimes$ denotes the Kronecker product, and the operator `$\operatorname{vec}$' means vectorization.

The whole procedure for solving $\mathcal{P}\text{(C)}$ is summarized in Algorithm~\ref{algorithm:case3}.
\begin{algorithm}[t] 
\caption{The Penalty Method for Solving $\mathcal{P}\text {(C)}$ } 
\begin{algorithmic}[1] \label{algorithm:case3}
\STATE Initialize the $\{\mathbf{t}_n\}_{n=1}^N$, $\{\mathbf{u}_n\}_{n=1}^N$, and $\mathbf{W}$.
\REPEAT
\STATE Optimize $\mathbf{W}$ according to~\eqref{optimal_W}.
\STATE \textbf{for} $n=1,\ldots,N$ \textbf{do}
\STATE ~~Optimize $\{\mathbf{t}_n\}_{n=1}^N$ with iteration on~\eqref{pgm_t_case3} with respect to $i$.
\STATE \textbf{end for}
\STATE \textbf{for} $n=1,\ldots,N$ \textbf{do}
\STATE ~~~~Optimize $\mathbf{u}_n$ using Algorithm~\ref{a2}, with input $\mathbf{r}_m$ replaced by $\mathbf{t}_n$ and $\mathbf{z}_l$ replaced by $\mathbf{u}_l$.
\STATE \textbf{end for}
\STATE Increase the penalty factor $\rho$.
\UNTIL Stopping criterion is satisfied.	
\end{algorithmic}
\end{algorithm}



\section{Numerical Results}
\label{Numerical Results}

In this section, we demonstrate the superiority of the proposed optimization framework on three cases introduced in Section~\ref{case}. Unless specified otherwise, all parameters used in this section are set as follows. The stopping criteria of the projected gradient descent of $\mathbf{r}_m$ and/or $\mathbf{t}_n$ and that of the outer loop in Algorithms~\ref{algorithm:case1} to~\ref{algorithm:case3} are set as the relative change of the respective penalized objective function values between two adjacent iterations being smaller than $10^{-3}$. In all cases, we consider the transmitter and receiver are equipped with $N=6$ and $M=6$ MAs, respectively. The given regions for MA, i.e., $\mathcal{C}_{\text{R}}$ and $\mathcal{C}_{\text{T}}$, are set as $A \times A$ square areas, which corresponds to $\mathbf{x}_{\text{min}} = \left[-A/2, -A/2 \right]^T$ and $\mathbf{x}_{\text{max}} = \left[A/2, A/2\right]^T$, respectively. The minimum distance between MAs for avoiding the coupling effect is $D=\lambda/2$. All simulations are based on the geometric channel model~\cite{MIMO_cap_cha_for_MA}, which assumes that the number of transmit paths is the same as that of receive paths. Thus, in this section, $L_t$ and $L_r$ will be replaced by $L$ for simplicity, and we use $L = 10$. It is also assumed that the path response matrix is a diagonal matrix whose non-zero entries are given by $\mathbf{\Sigma}[1,1] \sim \mathcal{CN}(0, \frac{\kappa}{(\kappa+1)})$ and $\mathbf{\Sigma}[l,l] \sim \mathcal{CN}(0, \frac{1}{(\kappa+1)(L-1)})$ for $\forall l \in \{2,3,\dots,L\}$, where $\kappa$ represents the ratio of the average power for line-of-sight paths to that for non-line-of-sight paths, and we set $\kappa = 1$. The elevation and azimuth angles are assumed to be i.i.d. variables uniformly distributed over $[0,\pi)$. The SNR is represented by $P/\sigma^2 = 15 \text{dB}$, where the noise is assumed to be of unit power. All the results are obtained through Monte-Carlo simulations on Matlab-R2023b, and each point in the figures is obtained by averaging over $500$ trials. The penalty factor $\rho$ is initialized as $5$ and increased by a factor of $1.2$ after each outer iteration.

\subsection{Capacity Maximization for MA-Aided MIMO System}

\begin{figure*}  
	\centering
    \subfigure[The convergence behavior of proposed framework]{\label{CoM_convergence} 
		\includegraphics[width=0.25\linewidth]{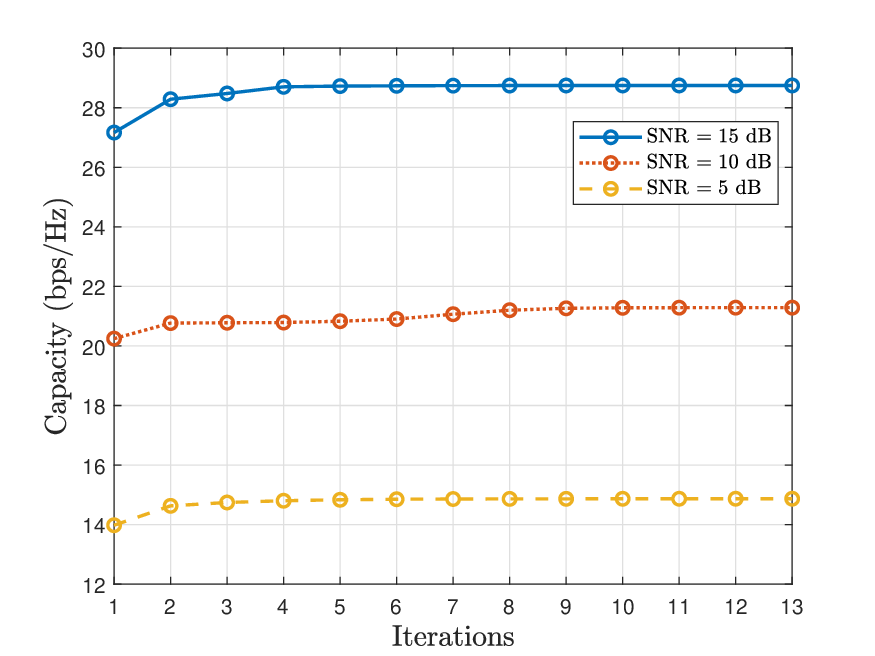}} 
    \subfigure[The capacity versus the normalized transmit/receive region size]{\label{CoM_performance} 
		\includegraphics[width=0.25\linewidth]{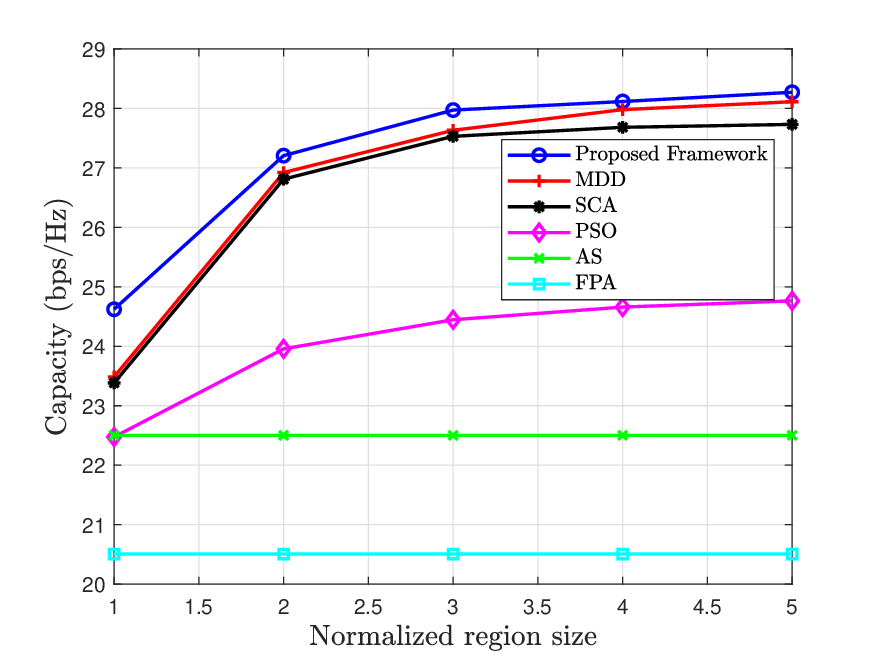}} 
	\subfigure[The capacity versus the number of MAs]{\label{CoM_4to12} 
		\includegraphics[width=0.25\linewidth]{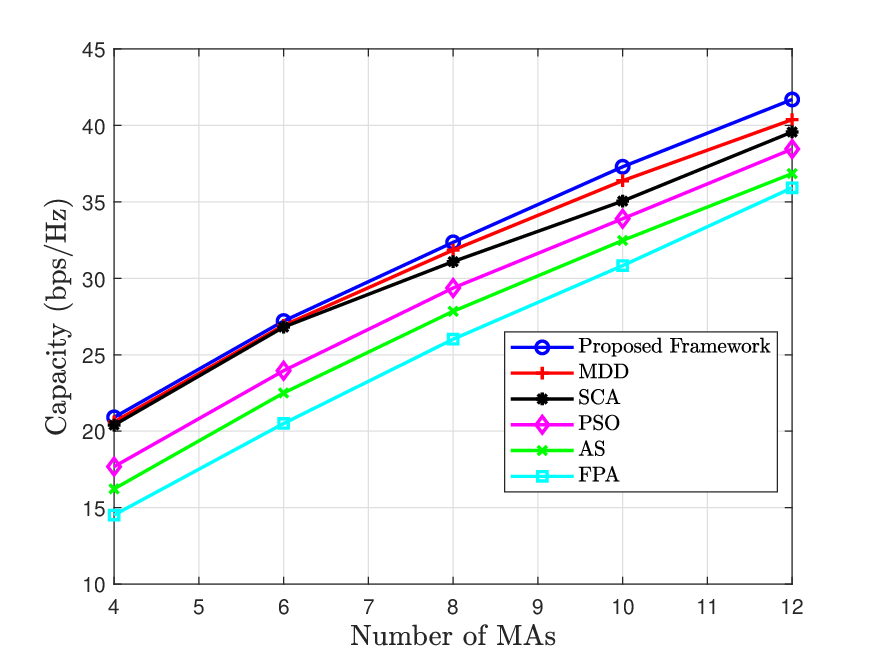}}  
        
  	\caption{Numerical results of case study 1: Capacity Maximization for MA-Aided MIMO System.} 
    \label{case1}
\end{figure*}

Firstly, we demonstrate the convergence of the proposed framework under the setting $A = 5\lambda$. The convergence behavior is shown in Fig.~\ref{CoM_convergence} for three different SNRs. The simulation results show that the achievable channel capacity monotonically increases and converges to stable values within $13$ outer iterations. The final value of $\rho$ is approximately $44.58$, while the objective function value of $\mathcal{P} \text{1-b}$ is $0$. This implies that while the theory of the penalty framework requires $\rho$ to be infinity to ensure the MA positions and the auxiliary variables be identical, this happens even under a finite and not very large $\rho$.


Next, the proposed framework is compared with the following state-of-the-art methods:
\begin{itemize}
    \item FPA: The $6$ antennas at both the BS and user device are spaced by $\lambda/2$, and are not optimized or changed with the size of the antenna panel. The capacity of this FPA system is obtained by the proposed algorithm without optimizing the position of antennas.
    
    \item AS: In this scheme, both the transmitter and receiver are equipped with $12$ fixed antenna candidates, spaced by $\lambda/2$, and the AS scheme selects half of the transmit and receive antennas to maximize the objective function via exhaustive search.

    \item PSO~\cite{FA_MEC}: The positions of the $6$ MAs in transmitter and receiver are updated by the particle swarm optimization (PSO) with a penalty function.

    \item Multi-directional descent (MDD)~\cite{MA_Multiusercomm_APO}: The positions of the $6$ MAs in the transmitter and receiver are updated along different descent directions in each iteration. With multiple update results, MDD selects the best solution among them as the final result of this update.
    
    \item SCA~\cite{MIMO_cap_cha_for_MA}: The $6$ MAs in both the transmitter and receiver are alternately optimized with the transmit covariance matrix by leveraging SCA.
\end{itemize}

In Fig.~\ref{CoM_performance}, we demonstrate the channel capacity versus the normalized region size $A/\lambda$ for the proposed optimization framework and other methods. The results show that the proposed optimization framework outperforms all baselines. Specifically, compared with FPA and AS, the proposed optimization framework and other algorithms with MAs can flexibly optimize the positions of the antennas in a given continuous region, thus enabling a higher spatial DoF for enhancing the channel capacity. 

Compared with the SCA method in~\cite{MIMO_cap_cha_for_MA}, the proposed framework achieves better performance as it involves no approximation. Compared with the PSO algorithm, the proposed framework exhibits significant superiority in performance. In practice, zeroth-order algorithms such as PSO often have a large number of hyperparameters to adjust. Finding the optimal settings for such algorithms can be a challenging task. Furthermore, although the MDD method can provide nearly identical capacity in the scenario of large antenna panel region, it divides the feasible set into multiple regions to satisfy the antenna distance constraints, which can strongly degrade the optimization performance when the antenna panel region is small or the number of MA is large. Therefore, when it is necessary to deploy a large number of MAs compactly, the proposed framework has consistent advantages over other methods.

Next, we demonstrate the performance of the proposed framework and other baselines versus different numbers of MAs. The length of the square antenna panel area is set as $A = 2\lambda$ in this simulation. The numbers of MAs at both the user device and the BS are set as the same number.
In Fig.~\ref{CoM_4to12}, it is shown that under the same number of MAs, the proposed framework consistently outperforms other methods, and this is consistent with Fig.~\ref{CoM_performance}. An important phenomenon worth noting is that, as the number of MAs increases, the given region becomes more crowded, and the objective function of the optimization problem gradually becomes more complex. Therefore, in scenarios where the number of MAs is small, the performance gap between the SCA, MDD and the proposed framework is small. However, as the number of MAs increases, the effectiveness of the SCA method in approximating the objective function compromises, and the feasible set loss of the MDD algorithm gradually becomes larger. Consequently, the performance gap between the two baselines and the proposed framework consistently increases.

\subsection{Latency Minimization for MA-Aided MEC}

This case study focuses on an MEC system consisting of $6$ single-FPA users and a BS equipped with MAs. The bandwidth $b$ is $100$ MHz. Futhermore, for easy comparison with the results of~\cite{FA_MEC}, it is assumed that all users are arranged along a straight line, with coordinate $\mathbf{t}_n = \left[(n-1)d,0 \right]^T$ for $n= 1,2,\dots,N$, and $d$ refers to the distance between adjacent mobile users.

\begin{figure*}  
	\centering
    \subfigure[The convergence behavior of proposed framework]{\label{MAMEC_convergence} 
		\includegraphics[width=0.25\linewidth]{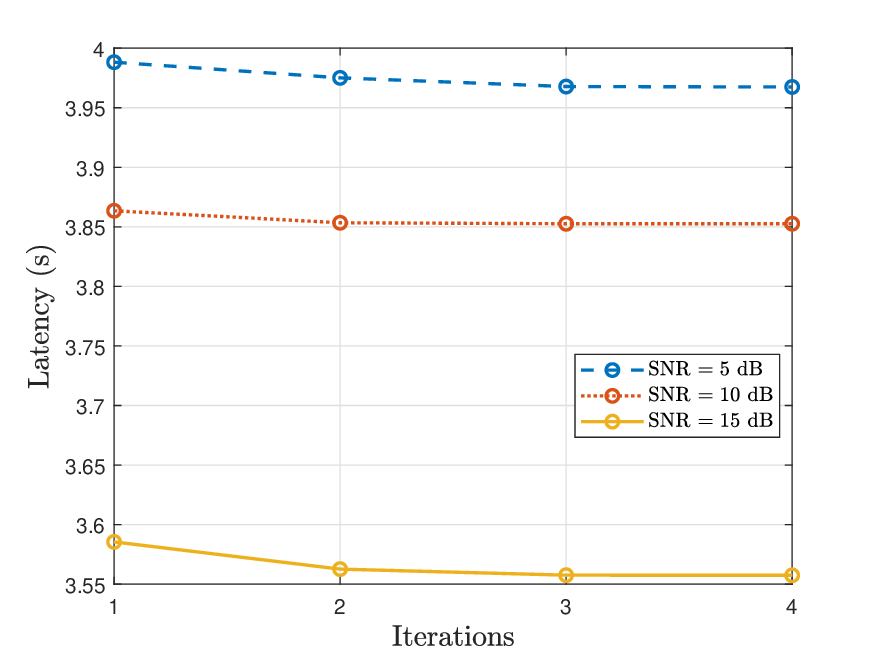}} 
    \subfigure[The latency versus the normalized antenna panel region size]{\label{MAMEC_performance} 
		\includegraphics[width=0.25\linewidth]{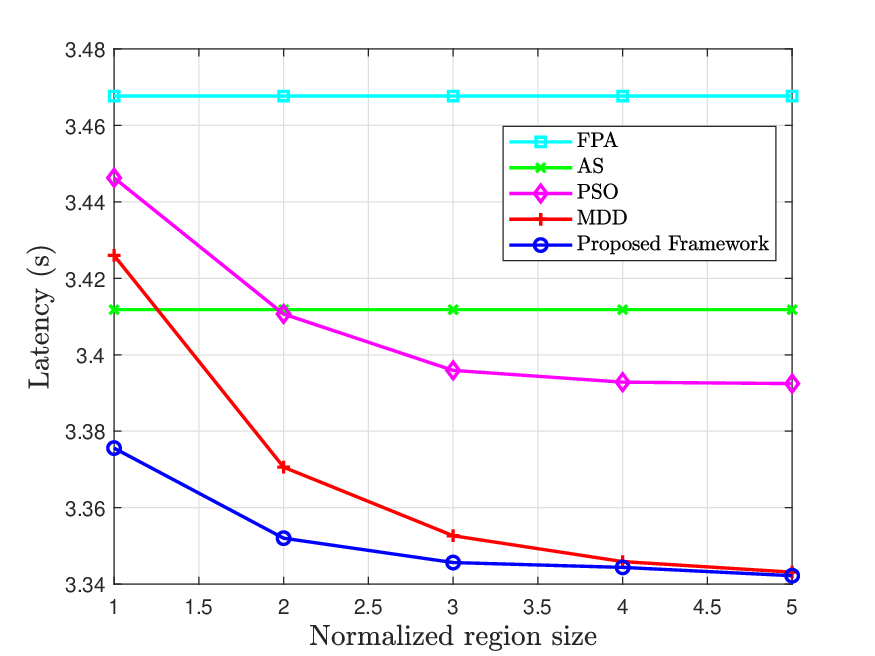}} 
	\subfigure[The latency versus the number of receive MAs]{\label{MAMEC_4to12} 
		\includegraphics[width=0.25\linewidth]{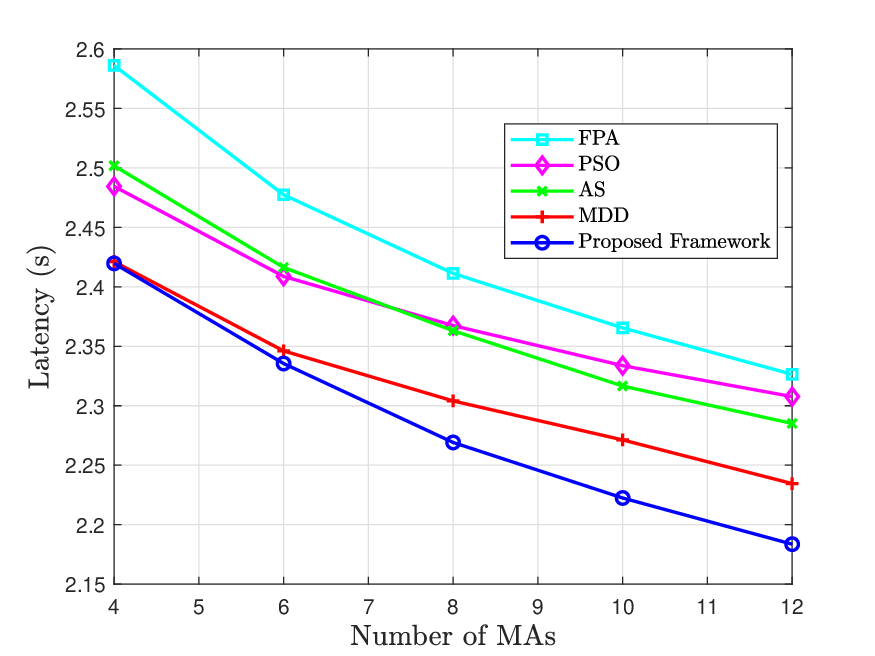}}  
        
  	\caption{Numerical results of case study 2: MA-Aided Mobile Edge Computing.} 
    \label{case2}
\end{figure*}

First, we demonstrate the convergence of the proposed framework under the setting $A = 5\lambda$ in Fig.~\ref{MAMEC_convergence}. It can be seen that the network latency monotonically decreases and converges to the stable values within $4$ outer iterations, validating the convergence of the objective function value. The final value of $\rho$ is approximately $8.64$, while the objective function value of $\mathcal{P} \text{1-b}$ is $0$. This again shows that we do not need a very large $\rho$ to make the optimized MA positions to be identical to the auxiliary variables, and thereby satisfy the antenna distance constraints.


Then, we compare the performance of the proposed framework with state-of-the-art methods under different normalized receive region size $A/\lambda$.
It is evident from Fig.~\ref{MAMEC_performance} that the performance of the proposed framework is superior to all benchmarks in this MA-aided MEC scenario. Additionally, compared with the MDD method, the proposed framework performs better under the condition of small antenna panel region.

Next, we explore the relationship between the latency of the MEC network and the number of MAs at the receiver. The number of transmitters is $N = 4$, and $A = 2\lambda$. 
As shown in Fig.~\ref{MAMEC_4to12}, the proposed framework provides solutions with superior quality compared to other benchmarks. Additionally, as the number of receive antennas increases, the advantages of proposed framework become increasingly apparent.

\subsection{Regularized Zero-Forcing Precoding for MA-Aided Multi-User System}

\begin{figure*}  
	\centering
    \subfigure[The convergence behavior of proposed framework]{\label{FP_convergence} 
		\includegraphics[width=0.25\linewidth]{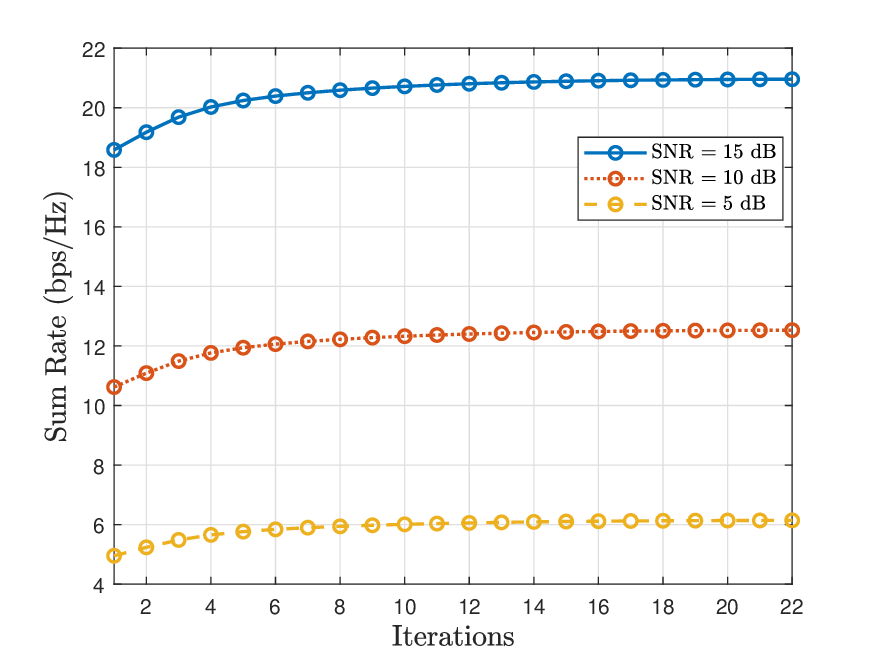}} 
    \subfigure[The sum rate versus the normalized transmit antenna panel size]{\label{FP_performance} 
		\includegraphics[width=0.25\linewidth]{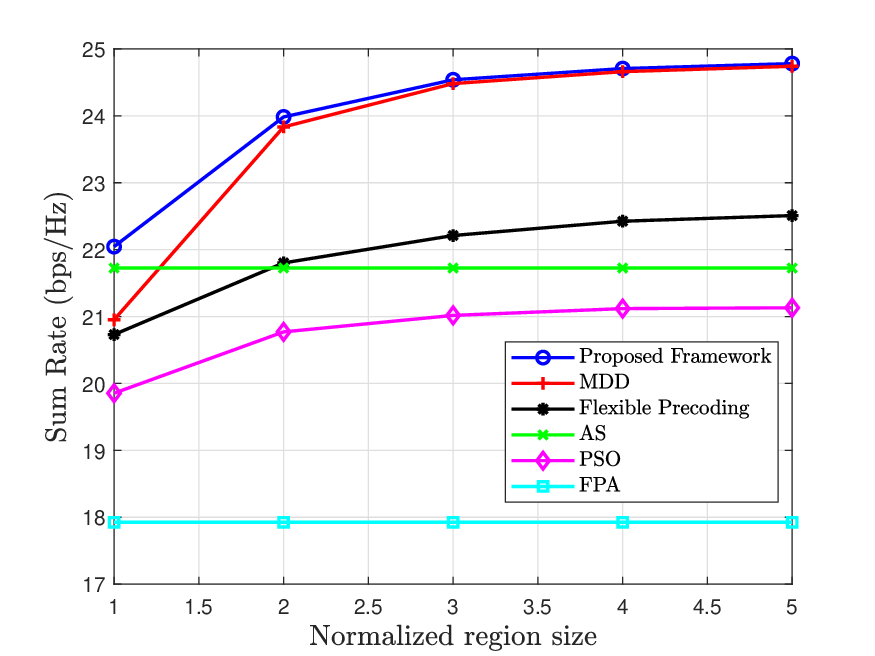}} 
	\subfigure[The sum rate versus the number of transmit MAs]{\label{FP_4to12} 
		\includegraphics[width=0.25\linewidth]{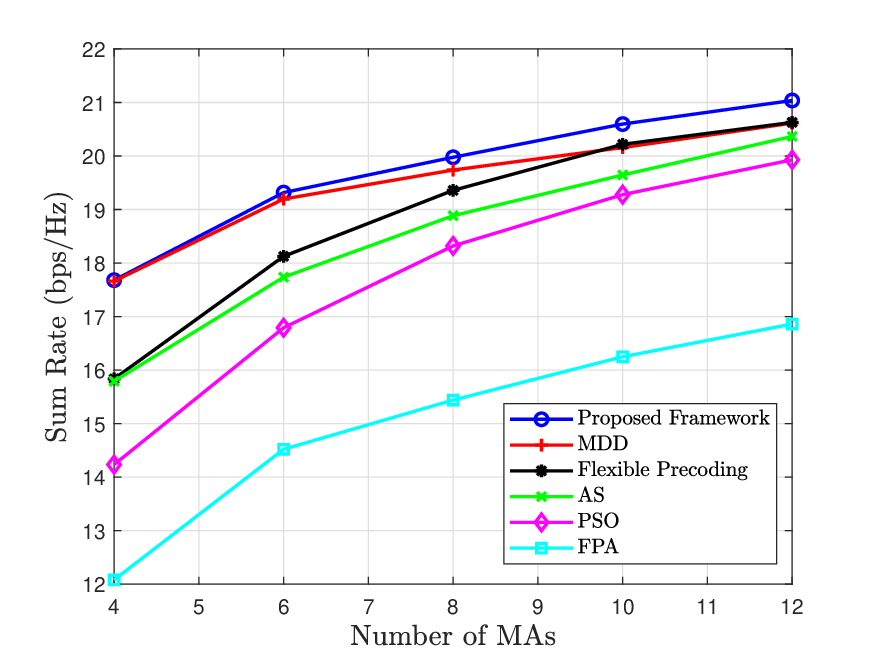}}  
        
  	\caption{Numerical results of case study 3: Regularized Zero-forcing
    Precoding for MA-Aided Multi-user System.} 
    \label{case3}
\end{figure*}

In this case, we consider the multi-user MISO system with $6$ single-FPA users (positions of users are arranged as in~\cite{MIMO_cap_cha_for_MA}), and the BS employs MAs to enhance the performance of downlink transmission. First, we demonstrate the convergence of the proposed framework in solving this precoding problem in Fig.~\ref{FP_convergence}. The results indicate that the sum rate of this system monotonically increases and converges to a stable value within $22$ outer iterations. The final value of $\rho$ is approximately $230.03$, while the objective function value of $\mathcal{P} \text{1-b}$ is $0$. This confirms the practicality of the proposed penalty framework to give useful solution even under a small $\rho$. 


In addition to the baselines in the previous cases, we also compare with the following scheme which is specially designed for this case study:
\begin{itemize}
    \item Flexible Precoding (FP)~\cite{Flexible_precoding_MU_MAcomm}: This method incorporates a zero-norm constraint and utilizes CS-based approach that combines the ideas of sparse optimization (SO) and subspace projection to design the precoder for MA-aided systems.
\end{itemize}

Fig.~\ref{FP_performance} illustrates the sum rate versus the normalized transmit antenna panel size, $A/\lambda$. As demonstrated in this figure, the performance of proposed framework surpasses all baselines. The gap between the proposed framework and the FP method can be attributed to the avoidance of conservative approximations. More importantly, in contrast to the FP algorithm which was tailored to a specific problem, the proposed framework exhibits a broader applicability.

Finally, Fig.~\ref{FP_4to12} explores the relationship between the sum rate of the system and the number of transmit MAs, $N$. This simulated system involves $M = 4$ users, with the transmit antenna panel length $A = 2\lambda$. Fig.~\ref{FP_4to12} demonstrates that the FP method shows inferior performance when the number of antennas is restricted, whereas the MDD method faces challenges in scenarios with a large number of MAs due to the loss induced by conservative approximation of the feasible set. In contrast, the proposed framework consistently performs the best over a wide range for the number of MAs.


\section{Conclusion}
\label{Conclusion}

In this paper, a general resource allocation problem for MA-aided wireless systems has been studied. After revealing the challenges for handling the non-convex and coupled antenna distance constraints by existing methods, we have introduced auxiliary variables to separate the non-convex constraints from the objective function, and have proposed a penalty optimization framework to solve the problem. The transformed problem was then solved via alternating optimization, with the solutions of the subproblem with respect to the auxiliary variables obtained in closed-form. To demonstrate the superiority of the proposed framework, we applied it to three case studies: capacity maximization, latency minimization, and RZF precoding. Numerical results confirmed the superior performance of proposed framework compared to other state-of-the-art methods. It is expected that the results of this paper would facilitate the incorporation of MA to other wireless technologies.

\appendices

\section{Proof of Proposition~\ref{lemma: union of u&w and optimal z}}
\label{appendix: optimal solution}

First, we consider the case of $|\mathcal{L}|=1$. By the definition of $\mathcal{L}$, $|\mathcal{L}|=1$ means $\mathbf{r}_m$ must lie within exactly one $\text{Circle}_{l}$.  More precisely, if the element in $|\mathcal{L}|$ is denoted as $*$, $\mathbf{r}_m$ lies within $\text{Circle}_*$ (e.g., $m=1$ and $*$ equals $4$ in the illustration of Figs.~\ref{222} and ~\ref{333}). We claim that the optimal solution must be on the boundary of $\text{Circle}_*$. This is because according to the constraints of $\mathcal{P}\text{1-b-m}$, they mandate that the solution cannot be situated within $\text{Circle}_*$. On the other hand, if a solution $\mathbf{o}$ were to lie outside $\text{Circle}_*$, there always exists an intersection point $\mathbf{o}^\prime$ between the line formed by $\mathbf{r}_m$ and $\mathbf{o}$ and the $\text{Circle}_*$ such that $||\mathbf{o}^\prime - \mathbf{r}_m||_2 < ||\mathbf{o} - \mathbf{r}_m||_2$. Therefore, the optimal solution of $\mathcal{P}\text{1-b-m}$ must be on $\text{Circle}_*$. 

\begin{figure}
    \centering
    \includegraphics[width=0.35\linewidth]{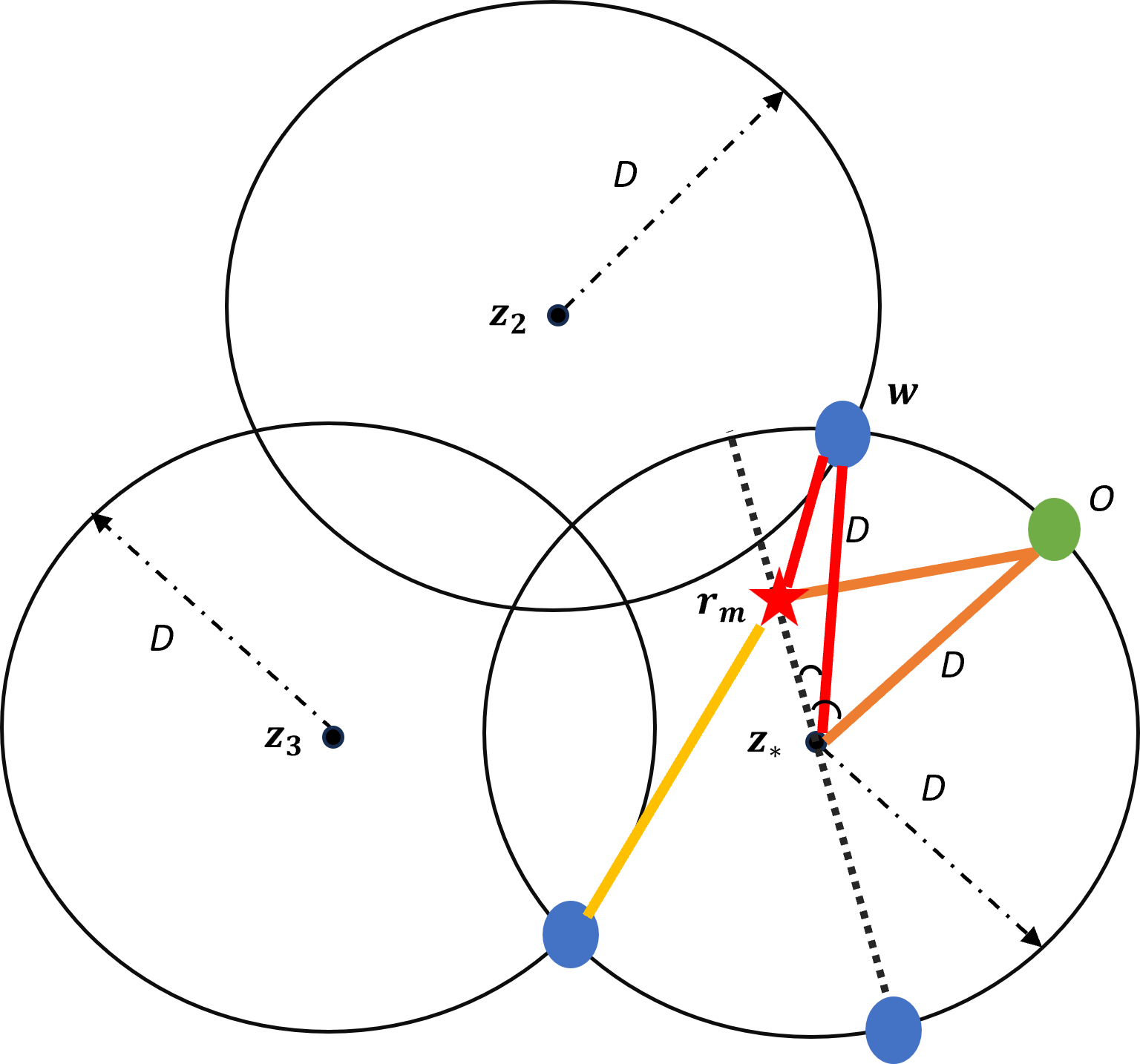}
    \caption{Illustration for the proof of Proposition~\ref{lemma: union of u&w and optimal z}.}
    \label{fig: figure for appendix}
\end{figure}

Next, we prove that the optimal solution must be in the set $\mathcal{W}_* \bigcup \mathcal{U}_*$, which are points on $\text{Circle}_*$.  
\begin{itemize}
    \item Suppose $\mathcal{U}_*$, which contains the intersection points on $\text{Circle}_*$ with the line passing through $\mathbf{r}_m$ and $\mathbf{z}_*$, has two elements. Then one of the elements in $\mathcal{U}_*$ must be the optimal solution of $\mathcal{P}\text{1-b-m}$ as it has the smallest distance from $\mathbf{r}_m$ among all the points on $\text{Circle}_*$. An example is shown in Fig.~\ref{222}, with the upper orange point, which is an element in $\mathcal{U}_4$, being the optimal solution.

    \item However, it may be possible that one or both points in $\mathcal{U}_*$ do not satisfy the distance constraints of $\mathcal{P}\text{1-b-m}$, making them not qualified as solution. Then we need to seek for solutions from other points on $\text{Circle}_*$. An example is shown in Fig.~\ref{333}. In this case, we claim that the optimal solution must be in $\mathcal{W}_*$. More specifically, referring to Fig.~\ref{fig: figure for appendix}, there are two triangles with vertices at $\mathbf{w}$, $\mathbf{r}_{m}$, $\mathbf{z}_*$, and $\mathbf{o}$, $\mathbf{r}_{m}$, $\mathbf{z}_*$, respectively. $\mathbf{r}_{m}\mathbf{z}_*$ is a common side of these triangles, while we are given that $||\mathbf{w} - \mathbf{z}_*||_2 = ||\mathbf{o} - \mathbf{z}_*||_2 = D$. According to the Law of Cosines, we have $||\mathbf{r}_{m} - \mathbf{w}||_2 = \sqrt{a^2~+~||\mathbf{w} - \mathbf{z}_*||_2^2~-~2a \cdot ||\mathbf{w} - \mathbf{z}_*||_2~ \cos{\angle\mathbf{r}_{m}\mathbf{z}_* \mathbf{w}}}$, and $||\mathbf{r}_{m} - \mathbf{o}||_2 = \sqrt{a^2~+~||\mathbf{o} - \mathbf{z}_*||_2^2~-~2a~\cdot~ ||\mathbf{o} - \mathbf{z}_*||_2 ~\cos{\angle\mathbf{r}_{m}\mathbf{z}_*\mathbf{o}}}$, where $a = ||\mathbf{r}_{m} - \mathbf{z}_*||_2$. Considering that $\angle\mathbf{r}_{m} \mathbf{z}_*\mathbf{w}< \angle \mathbf{r}_{m}\mathbf{z}_*\mathbf{o}$, it is apparent that $||\mathbf{r}_{m} - \mathbf{w}||_2 < ||\mathbf{r}_{m} - \mathbf{o}||_2$. This shows that $\mathbf{w}$, which is in $\mathcal{W}_*$, is a better solution than $\mathbf{o}$ which is not in $\mathcal{W}_*$.
\end{itemize}

Based on the proof provided above, when $\mathbf{o} \notin \mathcal{W}_* \bigcup \mathcal{U}_*$, there will always be another point in the set $\mathcal{W}_* \bigcup \mathcal{U}_*$ that yields a superior objective function value of $\mathcal{P}\text{1-b-m}$. Consequently, the optimal solution for $\mathcal{P}\text{1-b-m}$ resides within the set $\mathcal{W}_* \bigcup \mathcal{U}_*$. Moreover, upon obtaining the set $\mathcal{W}_* \bigcup \mathcal{U}_*$, the optimal $\mathbf{z}_m^{\star}$ is determined by solving $\mathbf{z}_m^{\star} = \arg \min_{\mathbf{z}_m \in \mathcal{W}_* \bigcup \mathcal{U}_*} \left\|\mathbf{z}_m - \mathbf{r}_m\right\|_2^2$.

In conclusion, the lemma has been proved when $|\mathcal{L}|=1$. Similarly, it can be proved that this lemma still holds when $|\mathcal{L}| \geq 2$ by repeating the above proof for every element in $\mathcal{L}$.

\bibliographystyle{IEEEtran}
\bibliography{ref}
\vfill

\end{document}